\documentclass[final,5p,times,twocolumn]{elsarticle}
\usepackage{graphics}
\usepackage{amssymb}
\usepackage{url}
\usepackage{graphicx}
\usepackage{soul}
\usepackage{amsthm}
\usepackage{multicol}
\usepackage{textcomp}
\usepackage{booktabs}
\usepackage{amsmath}
\usepackage[draft, inline, nomargin]{fixme}
\fxsetup{theme=color, mode=multiuser}
\FXRegisterAuthor{ab}{1}{\color{blue}AB}
\usepackage{lineno}
\usepackage{threeparttable}
\usepackage[colorlinks]{hyperref}

\journal{Nuclear Instruments and Methods in Physics Research, Part A}

\begin{document}

\begin{frontmatter}

\title{SANDD: A directional antineutrino detector with segmented  \\ $^6$Li-doped pulse-shape-sensitive plastic scintillator }

\author[UM,LLNL]{F.~Sutanto}\ead{sutanto2@llnl.gov}
\author[LLNL]   {T.~M.~Classen}
\author[LLNL]   {S.~A.~Dazeley}\ead{dazeley2@llnl.gov}
\author[UH]     {M.~J.~Duvall} 
\author[UM]     {I.~Jovanovic} 
\author[LLNL]   {V.~A.~Li}
\author[LLNL]   {A.~N.~Mabe\footnotemark[1]}
\author[LLNL]   {E.~T.~E.~Reedy}
\author[UM]     {T.~Wu}

\address[UM]{Department of Nuclear Engineering and Radiological Sciences, University of Michigan, Ann Arbor, MI 48109}
\address[LLNL]{Lawrence Livermore National Laboratory, Livermore, CA 94550}
\address[UH]{Department of Physics and Astronomy, University of Hawai`i at M\={a}noa, Honolulu, HI 96822}

\begin{abstract}
\noindent We present a characterization of a small (9-liter) and mobile 0.1\% $^6$Li-doped pulse-shape-sensitive plastic scintillator antineutrino detector called SANDD (Segmented AntiNeutrino Directional Detector), constructed for the purpose of near-field reactor monitoring with sensitivity to antineutrino direction. 
SANDD comprises three different types of module. 
A detailed Monte Carlo simulation code was developed to match and validate the performance of each of the three modules. The combined model was then used to produce a prediction of the performance of the entire detector. 
Analysis cuts were established to isolate antineutrino inverse beta decay events while rejecting large fraction of backgrounds.
The neutron and positron detection efficiencies are estimated to be 34.8\% and 80.2\%, respectively, while the coincidence detection efficiency is estimated to be 71.7\%, resulting in inverse beta decay detection efficiency of 20.0\%$\pm$0.2\%(\textit{stat.})$\pm$2.1\%(\textit{syst.}). 
The predicted directional sensitivity of SANDD produces an uncertainty of 20$^\circ$ in the azimuthal direction per 100 detected antineutrino events.
\end{abstract}

\begin{keyword}
$^6$Li-doped plastic scintillator \sep pulse-shape discrimination \sep segmented scintillator \sep SiPM arrays \sep  aboveground reactor antineutrino detection 
\end{keyword}

\end{frontmatter}


\section{Introduction}

\footnotetext[1]{Deceased}

Nuclear power reactors are the most intense man-made sources of antineutrinos. 
A 1-GW thermal (GWt) nuclear reactor emits $\sim10^{20}$ antineutrinos per second isotropically. As the composition of the reactor core evolves due to the consumption of $^{235}$U and the production of $^{239}$Pu, the antineutrino flux and energy spectrum both evolve downwards ~\cite{Bowden:2006hu}. This makes it possible in principle to monitor the ON/OFF status and the fuel evolution of the core over time.
Antineutrinos have a very low interaction cross section, which makes their detection challenging. The same property allows them to pass through large amount of material unimpeded, which makes them unshieldable. 
The detection of antineutrinos presents opportunities for non-intrusive real-time monitoring of the operational status of nuclear reactors for safeguards, which falls within purview of the International Atomic Energy Agency~\cite{Bernstein2010,bernstein_rmp2020}. 
Additionally, short-baseline measurements of reactor antineutrinos could offer valuable insights into the nature of the neutrino~\cite{prospect2019,buck2017,Qian_2019,BOSER2019103736}.

The concept of continuous non-intrusive real-time monitoring of a reactor core using antineutrinos has been demonstrated several times. The concept was first demonstrated in the 1980's at the Rovno nuclear power plant in the Soviet Union~\cite{Korovkin1984,rovno1994}. It was followed by experiments at the Bugey nuclear power plant in France using $^6$Li-loaded liquid scintillator~\cite{CAVAIGNAC1984387,Bugey_direction} and in the 2000's at the San Onofre Nuclear Generating Station (SONGS) in the US~\cite{Bowden_ev} using a 0.64-ton Gd-doped liquid-scintillator detector deployed at an overburden of 20 meters water equivalent (m.w.e.). Recently, the Nucifer experiment, a 0.8-m$^3$ Gd-doped liquid scintillator detector located 7~m away from the 70-MWth Osiris reactor, provided a short baseline measurement of the reactor neutrino flux at an overburden of 12 m.w.e.~\cite{Boireau:2015dda}. 

Antineutrinos undergo the inverse beta decay (IBD) reaction on protons in a hydrogen-rich target volume:
\begin{equation}\label{Eq:ibd}
    \bar{\nu}_e + p \rightarrow e^{+} + n,
\end{equation}
producing an MeV-scale positron that slows down and annihilates with an electron, emitting two 0.511-MeV gamma rays. The neutron thermalizes and can be detected via capture on nuclei with high thermal neutron capture cross sections, usually added to the detection medium in small concentrations, such as $^{155}$Gd, $^{157}$Gd, and $^{6}$Li.

While the capabilities of antineutrino detectors for nuclear safeguards have been demonstrated, practicalities associated with deployment need to improve~\cite{Bernstein2010}. 
Most short-baseline detectors have been deployed below ground to suppress fast neutrons caused by the cosmic ray muons ~\cite{Bowden:2006hu,rovno1994}. Thick hydrogenous shielding is required around the target material to further reduce the fast neutron flux.
This increases the overall size and complexity of the detector and limits the range of possible deployment locations.
To solve this issue, contemporary near-field detectors employ segmented geometry and materials capable of pulse-shape discrimination to reject fast neutron induced backgrounds and to select the delayed coincidence events that are confined both in time ($<$200~{\textmu}s) and space ($<$1~m).
This approach greatly suppresses the accidental coincidence and the muogenic backgrounds, allowing for aboveground deployment. 
For example, PROSPECT is an aboveground segmented 4-ton $^6$Li-doped pulse-shape-sensitive liquid-scintillator detector that has demonstrated real-time monitoring of the 85-MW High-Flux Isotope Reactor (HFIR) at a 7-m standoff~\cite{prospect2018}. 

\begin{figure}[ht!]
	\begin{center}
    \includegraphics[width=0.4\textwidth]{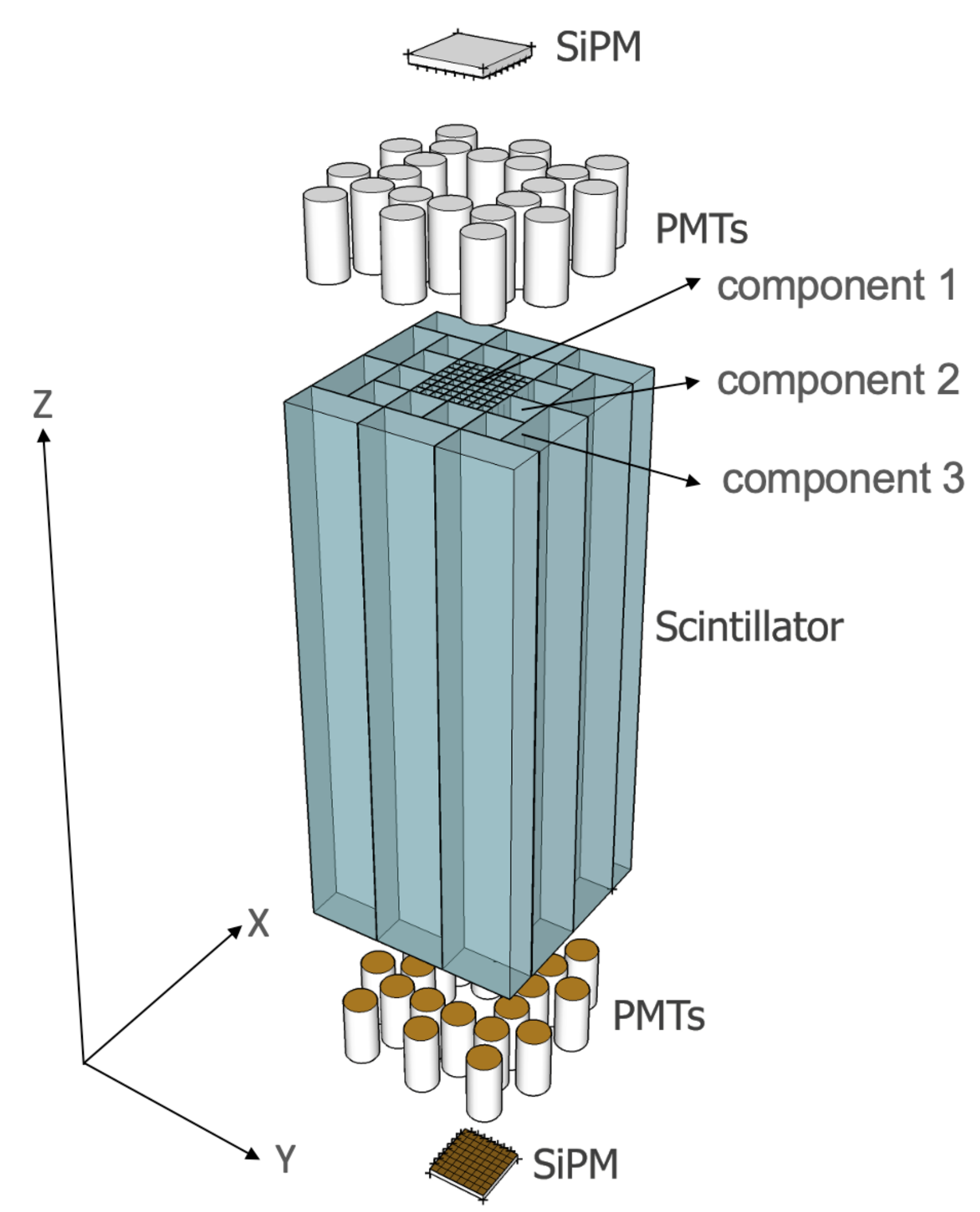}
	\end{center}
    \caption{Exploded view of the final configuration of the SANDD detector to be constructed and deployed near a nuclear reactor~\cite{SANDD1}. Here the central module, one 2.54~cm $\times$ 2.54~cm $\times$ 40.64~cm bar, and one 5.08~cm $\times$ 2.54~cm $\times$ 40.64~cm bar are referred to as component 1, component 2, and component 3, respectively.}\label{fig_SANDD_diagram}
\end{figure}

All of the short-baseline detectors described above
employed organic liquid scintillators. Some of these liquids can be chemically toxic, volatile, and flammable.
Energy depositions in the container holding the liquid may also not contribute to light production and detection.
Furthermore, materials compatibility with the liquid scintillator containment can be problematic, and support and retaining structures require careful engineering. To improve the safety characteristics and ease of deployment, several efforts to develop compact aboveground solid (plastic) antineutrino detectors are underway. Current and proposed near-field detectors such as SANDD~\cite{SANDD1,sutantoThesis}, PANDA~\cite{Panda2019}, SoLid~\cite{Abreu:2017bpe,Abreu:2018pxg}, DANSS~\cite{Alekseev:2013dmu,Alekseev:2016llm}, CHANDLER~\cite{chandler2019}, VIDAAR~\cite{vidaar}, IMSRAN~\cite{imsran}, and NuLat~\cite{Lane:2015alq,nulat2019,nulatThesis} are segmented and employ pulse-shape-sensitive solid materials.
These detectors \footnote{with an exception of NuLat and SANDD.} utilize heterogeneous detection medium: the main ``target'' scintillator is wrapped in neutron-capture material either containing Gadolinium (as in PANDA, DANSS, VIDAAR, and ISMRAN) or ${^6}$Li in the form of ${^6}$LiZnS:Ag (as in CHANDLER and SoLid).

Among the aforementioned detectors, SANDD (Fig.~\ref{fig_SANDD_diagram}) is the only detector that uses pulse-shape-sensitive plastic homogeneously doped with $^6$Li~\cite{ZAITSEVA2013747} (0.1\%wt).
In contrast to neutron capture on $^{155}$Gd, $^{157}$Gd, or H, which produces one or more gamma rays that deposit their energies over a larger volume, neutron capture on $^6$Li produces an alpha and a triton, which deposit their energies over a short range. 
Hence, the use of $^6$Li combined with the fine segmentation enables additional discrimination for/against neutron captures via rod multiplicity, improving IBD coincidence identification and further suppressing the high rate of backgrounds near the surface.
The fine lateral and azimuthal segmentation of the scintillator elements also permits good spatial resolution, a requirement for directional sensitivity. Antineutrino direction can be statistically reconstructed from the average displacement vector defined by the positron and neutron capture locations~\cite{Vogel:1999zy}.
Detectors such as Goesgen~\cite{goesgenPaloVerde_direction,Gosgen_direction}, Palo Verde~\cite{goesgenPaloVerde_direction,PaloVerde_direction}, CHOOZ~\cite{Chooz_direction}, Double CHOOZ~\cite{cadenThesis,langbrandtnerThesis,caden_hawaii}, and more recently PROSPECT~\cite{prospect2020} have theoretically and experimentally studied their directional sensitivity to antineutrino flux. So far, Double CHOOZ reported the lowest angular uncertainty of $\sim$43$^\circ$ from 100 detected IBD events~\cite{caden_hawaii}.

The focus of this paper is a 
characterization of the full 9-liter SANDD (Fig.~\ref{fig_rods_assembly}). It builds on earlier work~\cite{SANDD1}, which characterized a prototype of the central module with shorter pulse-shape-sensitive rods and no $^6$Li doping. 
The aim of the previous work was to test a data acquisition system capable of recording key performance metrics such as pulse-shape-parameter (PSP), energy ($E$), z-position resolution and particle ID via segment multiplicity ($R_m$). The relevant performance metrics in this work are similar, but also include 
PSP of the neutron capture on $^6$Li. Fig.~\ref{fig_SANDD_diagram} shows a rendering of the full sized 9-liter SANDD. The full 9-liter SANDD described here employs the same readout for the central module, and a dual-end readout of scintillator bars with 1" photomultiplier tubes (PMTs) for the outer  modules. 

The central module of SANDD consists of an 8 $\times$ 8 array of $^6$Li-doped pulse-shape-sensitive plastic rods (5.4~mm $\times$ 5.4~mm $\times$ 40.64~cm) coupled to a pair of 64-pixel SiPM arrays (Fig.~\ref{fig_rods_assembly}(a)). The central module of SANDD is surrounded by a layer of 2.54~cm $\times$ 2.54~cm $\times$ 40.64~cm $^6$Li-doped pulse-shape-sensitive plastic bars, which is further surrounded by a layer of 5.08~cm $\times$ 2.54~cm $\times$ 40.64~cm $^6$Li-doped pulse-shape-sensitive plastic bars. Each bar is coupled to a pair of 1"~PMTs (Fig.\ref{fig_rods_assembly}b). The performance characteristics of the central module, one of the middle layer bars, and one of the outer layer bars were measured and compared separately to a GEANT4 simulation~\cite{AGOSTINELLI2003250,ALLISON2016186}. Henceforth we refer to each of these components as component 1, component 2, and component 3, respectively, as described in Table~\ref{tab_nomen}.

\begin{table*}[ht!]
\caption{Nomenclature used throughout this paper.} 
\label{tab_nomen}
\centering
\begin{tabular}{|c|c|c|c|}
\hline
Component   & Disambiguation & Wrapped in teflon & Readout \\
\hline
\hline
1 (central module)  & 8 $\times$ 8 array of rods   5.4~mm $\times$ 5.4~mm $\times$ 40.64~cm & No & two 64-pixel SiPMs\\
2 (one of 12 first-layer rods)  & 2.54~cm $\times$ 2.54~cm $\times$ 40.64~cm & Yes & two 1" PMTs\\
3 (one of 10 second-layer rods)  & 5.08~cm $\times$ 2.54~cm $\times$ 40.64~cm & Yes & two 1" PMTs \\
\hline
\end{tabular}
\end{table*}

The three different components of SANDD are shown in Fig.~\ref{fig_rods_assembly} and parameterized in Table~\ref{tab_nomen}. 
Analysis cuts were established to retain the majority of IBD events while removing a large fraction of background events.
As a result, antineutrino detection efficiency and uncertainty in the direction of the reactor antineutrino flux are predicted by a tuned simulation.

\section{Design of SANDD}

\begin{figure*}[ht!]
	\begin{center}
    \includegraphics[width=1.0\textwidth]{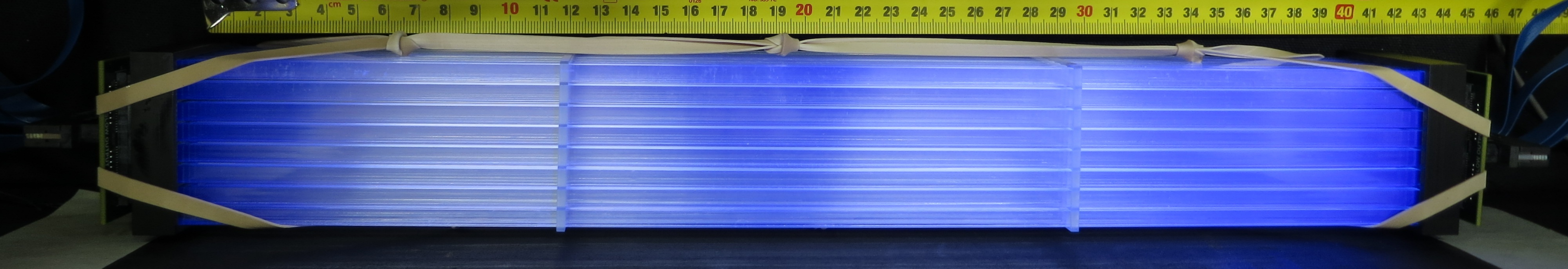}
(a)
    \includegraphics[width=1.0\textwidth]{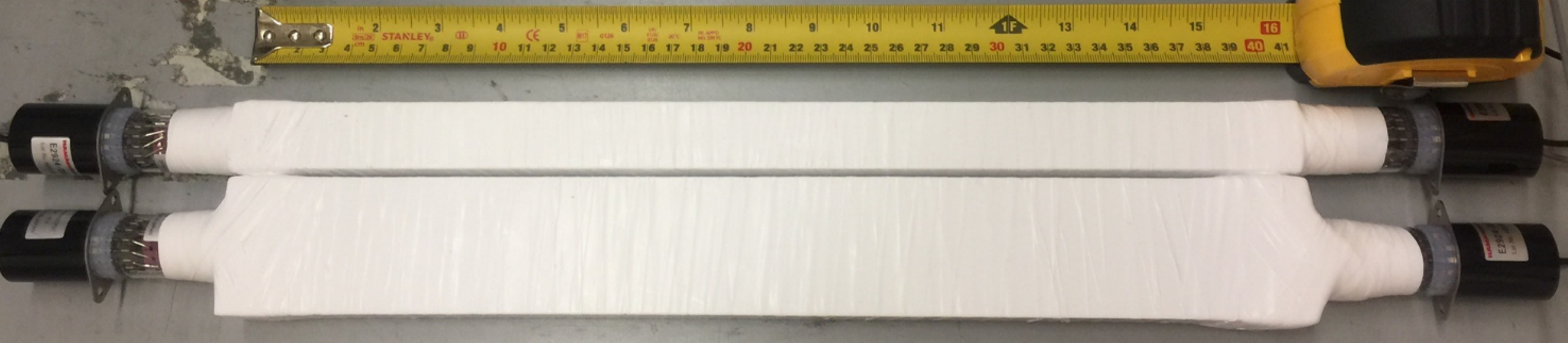}
(b)
    \end{center}
    \caption{SANDD consists of (a)~sixty-four $^6$Li-loaded plastic-scintillator rods, each of $0.54 \times 0.54 \times 40.6$~cm$^3$ dimensions, that are assembled into the $8 \times 8$-grid support frames and attached to the SiPM arrays (component 1) and (b)~larger cross section bars (2.54~cm $\times$ 2.54~cm $\times$ 40.6~cm and 5.08~cm $\times$ 2.54~cm $\times$ 40.6~cm) of the same material that are wrapped in PTFE tape and attached to 1" PMTs (component 2 and component 3).} \label{fig_rods_assembly}
\end{figure*}

Details of the composition and the fabrication of the $^6$Li-doped pulse-shape-sesitive plastic scintillator can be found in Ref.~\cite{Mabe_2019}. Multiple solid scintillator sections were prepared over an 8 months period. 
Some sections were machined for the 64-rod inner module, while the others were machined into the outer layer bars.
The rods were then optically coupled to a pair of SiPM arrays (SensL J-60035 series, 50.44~mm $\times$ 50.44mm~\cite{sensl_specSheet}) using BC-630 silicone optical grease, as shown in Fig.~\ref{fig_rods_assembly}(a). 
The scintillator rods were supported by an 8 $\times$ 8 square plastic frame that was fitted to each SiPM. The frames were used to maintain the alignment of each rod with its associated SiPM pixels.
Two additional 8$\times$8 square plastic frames were placed at one-third and two-thirds of the length of the rods to prevent them from sagging and contacting each other such that the total internal reflection within each rod can be maintained.
The larger bars were wrapped with polytetrafluoroethylene (PTFE) tape (POLY-TEMP PN-16050) and each end was optically coupled to a 1" Hamamatsu R1924A-100 PMT~\cite{hamamatsu_specSheet} using BC-630 silicone optical grease, as shown in Fig.~\ref{fig_rods_assembly}(b).
The light sensors (SiPMs or PMTs) mounted at either end of each plastic scintillator will be referred to as light sensors A and B.

The SiPM and PMT operating voltages were set at 28~V and $-$1100~V, respectively. The SiPM signals were amplified using CAEN N979 (10$\times$ gain) amplifiers while the PMT signals were sent directly to the Struck SIS3316 digitizer module (250~MS/s, 14~bit, 5~V dynamic range) without further amplification. Details of the data acquisition system can be found in Ref.~\cite{SANDD1}.
Thresholds were set to ensure data rates were not too high for the DAQ --- corresponding to an approximate energy threshold of 0.2~MeVee and 0.1~MeVee for the central module and the large cross section bars, respectively. Waveforms (1600~ns) were sent to disk as each digitizer buffer was filled.
The digitized waveforms were stored in ROOT format~\cite{Brun:1997pa}, and then processed by a series of standard ROOT routines.
In the analysis, individual channel timestamps were sorted into events consisting of waveforms within a 1~{\textmu}s event window. 
Energy depositions were required to trigger the light sensors on both ends of a given scintillator segment to be accepted for analysis.

\section{Characterization of the SANDD components and validation of the simulation framework}

\begin{figure*}[ht]
\begin{multicols}{3}
    \includegraphics[width=\linewidth]{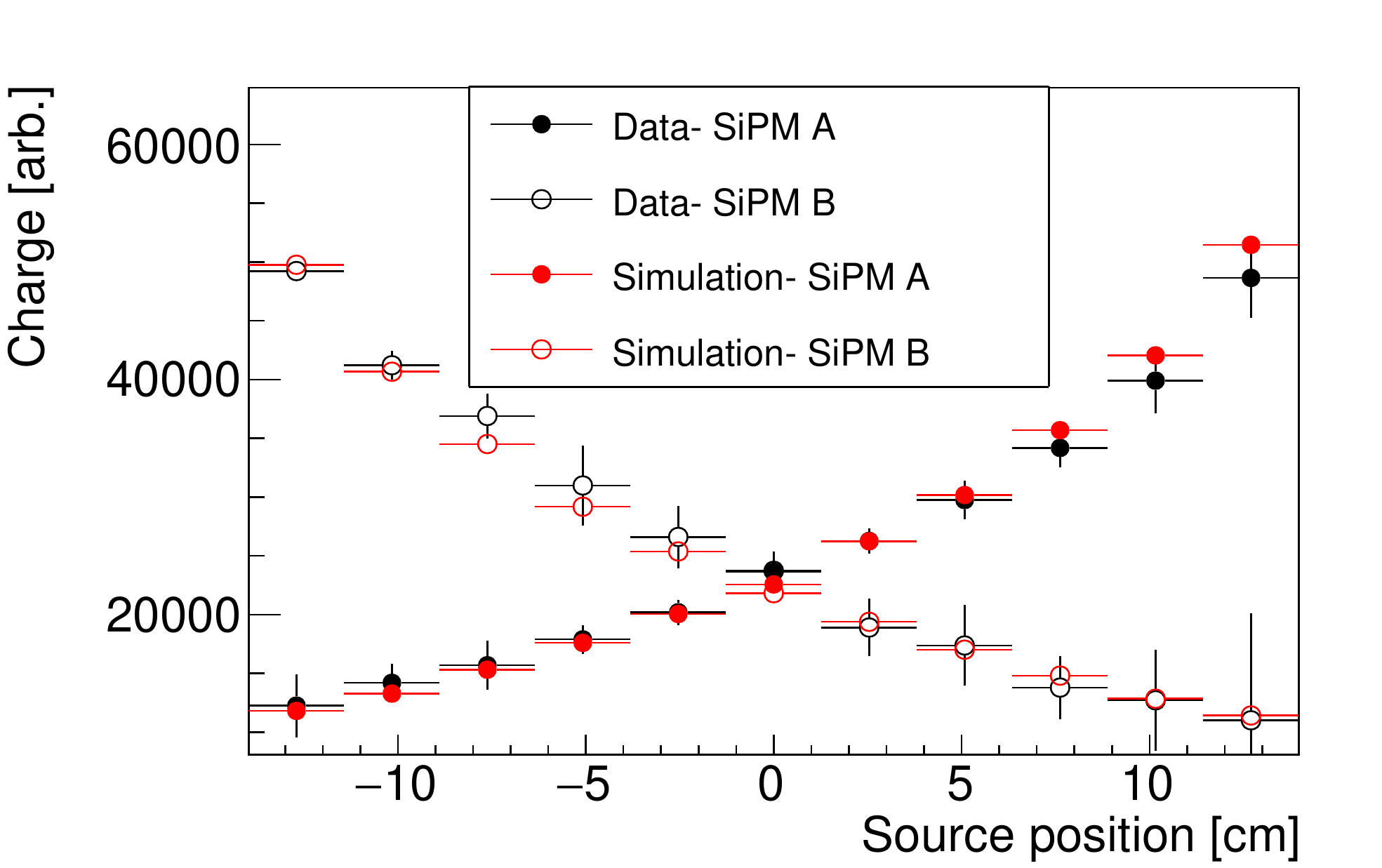} \par 
    \includegraphics[width=\linewidth]{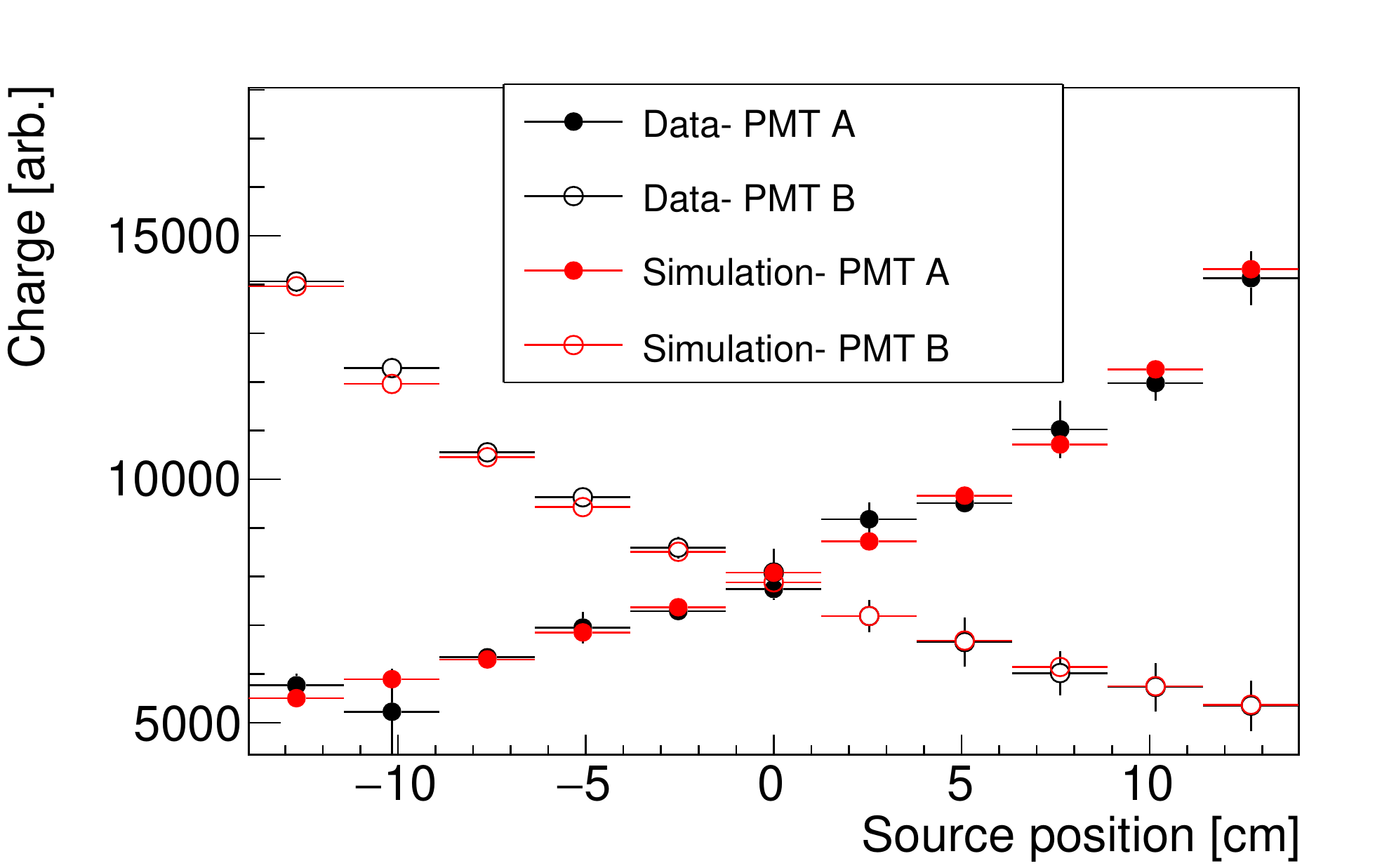} \par 
    \includegraphics[width=\linewidth]{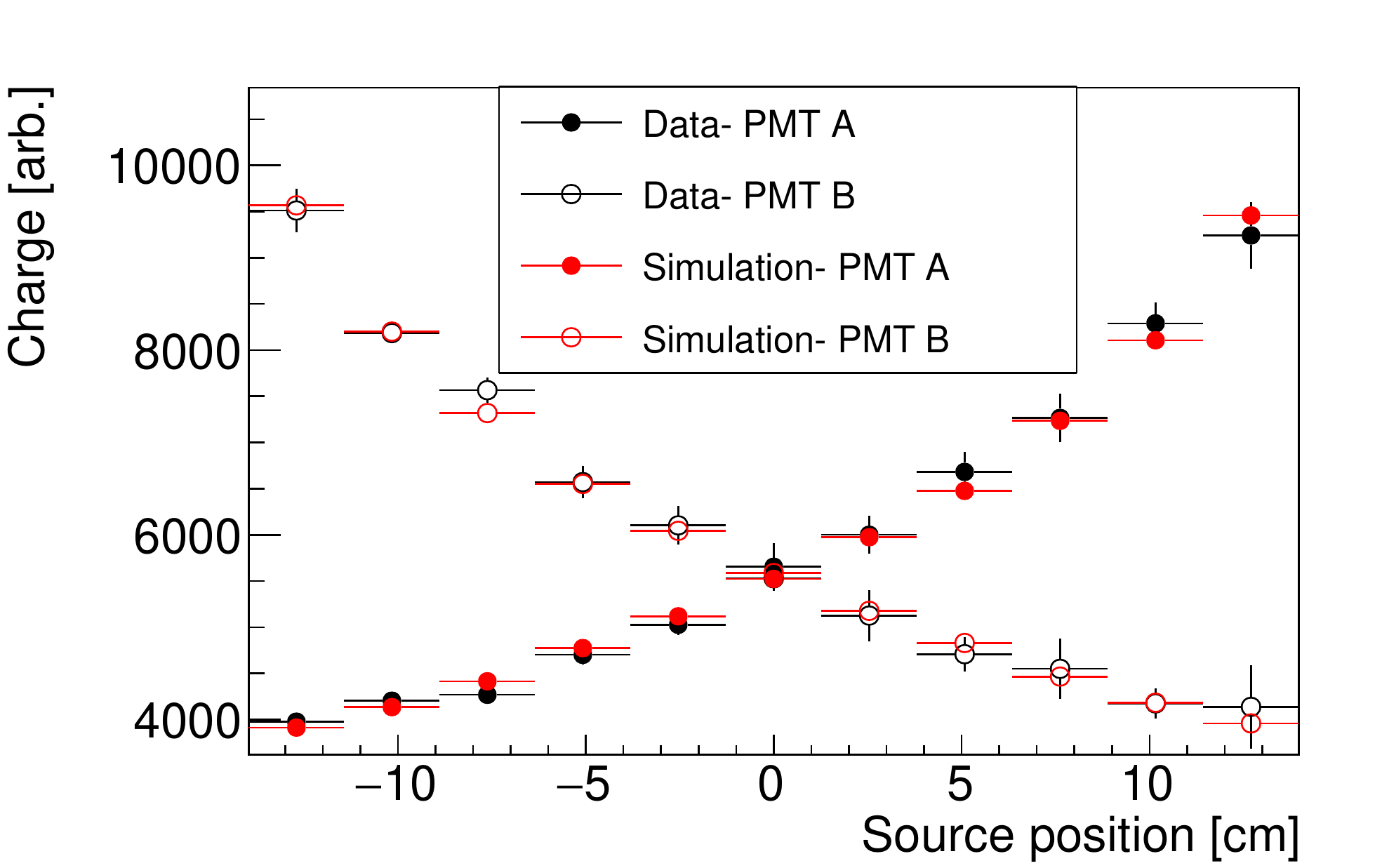} \par  
\end{multicols}
\begin{multicols}{3}
\begin{centering}
    (a) \par 
    (b) \par 
    (c) \par 
\end{centering}
\end{multicols}
\vspace{-0.3cm}
\caption{Comparison between the simulated and the measured attenuation length responses. The simulated responses were obtained with intrinsic attenuation length parameter of 72.6~cm and 
sigma alpha 
parameter of 0.12 for (a)~component 1, (b)~component 2, and (c)~component 3. The associated $\chi^2$/ndf value is 1.5.}
\label{fig_abs}
\end{figure*}

\begin{table*}[ht!]
\centering
 \begin{threeparttable}
\caption{Simulation parameters} 
\label{tab_param}
\centering
\begin{tabular}{|c||c c c| }
\hline
Parameters & Component 1 & Component 2 & Component 3 \\
\hline
\hline
sigma alpha   & & 0.12$\pm$0.02 & \\
Intrinsic attenuation length [cm] & & 72.6$\pm$10.3 & \\
$\chi^2$/ndf                      & & 1.5 (10~DOF)  & \\
\hline
\hline
Coupling efficiency$\times$Plastic yield [photons/MeV] &9103$\pm$24 &7867$\pm$1034 &7322$\pm$408 \\
$\chi^2$/ndf            &1.6 (20~DOF)    &1.6 (41~DOF)    &1.7 (41~DOF)          \\
\hline
\hline
Birks constant [mm/MeV] &0.22$\pm$0.004 &0.23$\pm$0.003 &0.24$\pm$0.004 \\
$\chi^2$/ndf            &1.9 (41~DOF)   &1.8 (23~DOF)   &1.9 (51~DOF)           \\
\hline
\hline
$^6$Li loading [\%wt] & &  0.1$\pm$0.01\tnote{a}   & \\
\hline
\end{tabular}
\begin{tablenotes}
      \item[a]{N. Zaitseva, personal communication, Aug 24, 2020.}
    \end{tablenotes}
\end{threeparttable}
\end{table*}

\begin{figure}[ht]
\centering\includegraphics[width=1.0\linewidth]{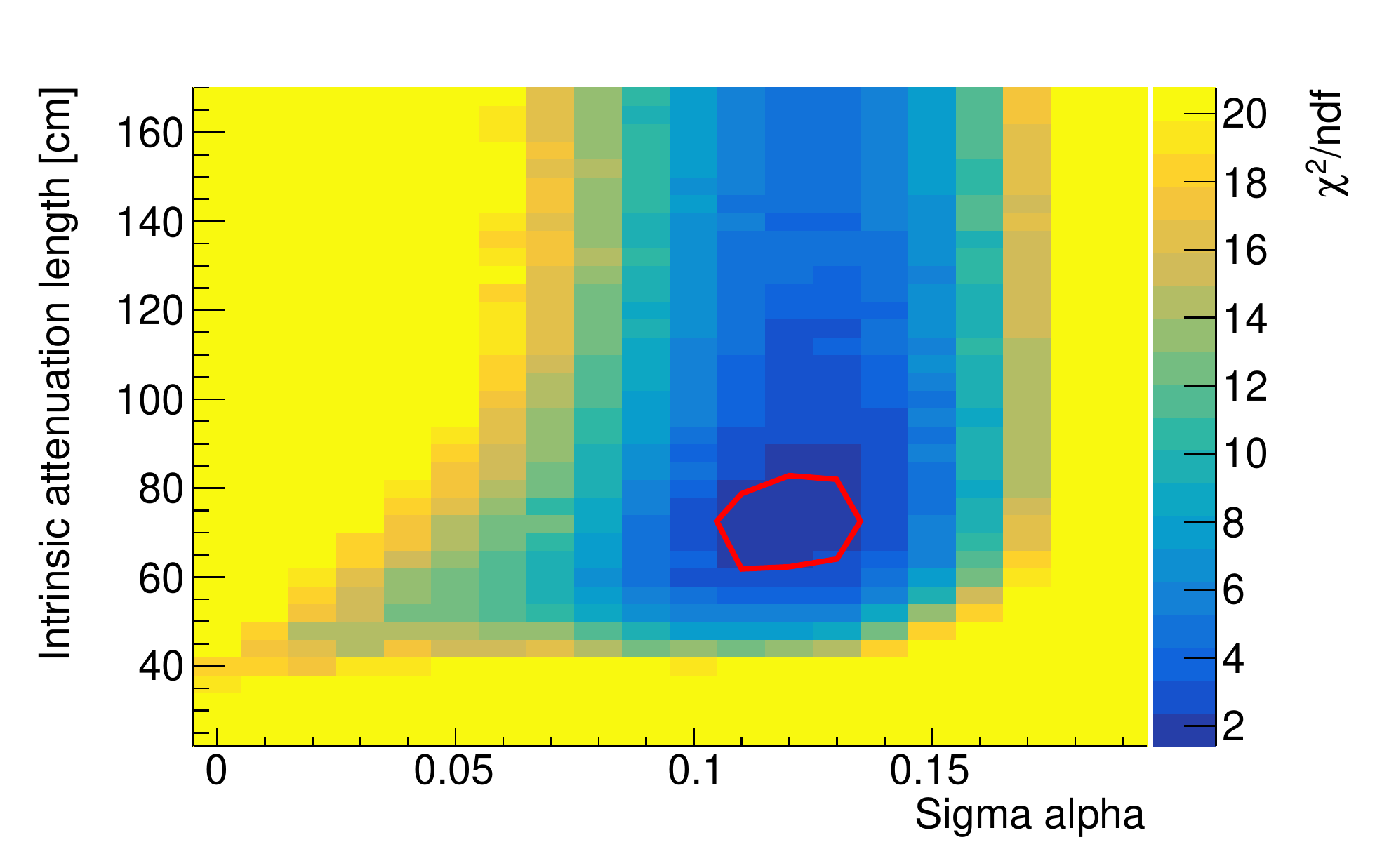}
\vspace{-0.3cm}
\caption{Distribution of goodness of fit ($\chi^2$/ndf) as a function of intrinsic attenuation length parameter and surface roughness (sigma alpha). The red contour indicates the 1-sigma region.}
\label{fig_chiAbs}
\end{figure}

%
%

The metrics that impact the performance of SANDD include energy resolution, 
PSP, light attenuation length (which impacts energy resolution), position resolution and rod multiplicity. A detailed description of the data analysis methods can be found in Ref.~\cite{SANDD1}. Here, only the pertinent aspects of those methods are highlighted. 

Waveforms saved to disk were analyzed as follows:
First, the baseline was calculated by averaging the first 70 pre-trigger samples (280~ns) and 30 pre-trigger samples (120~ns) of the digitized SiPM and PMT waveforms, respectively. 
The baseline was then subtracted on an event-by-event basis. 
The time of leading edge $t_{L}$ of each waveform was defined 
as the position of the maximum positive slope of the waveform.

Charge integration limits were optimized based on the PSP, where PSP is defined as
\begin{equation} \label{eq_psp}
    \mathrm{PSP} = \frac{Q_{\mathrm{tail}} }{ Q_{\mathrm{total}} } =
    \frac{\sqrt{ Q_{\mathrm{tail}}^{A\; }  Q_{\mathrm{tail}}^{B\; } } }
    {     \sqrt{ Q_{\mathrm{total}}^{A\;} Q_{\mathrm{total}}^{B\; }} }.
\end{equation}

Here, $Q_{\mathrm{tail}}^{A}$ ($Q_{\mathrm{tail}}^{B}$)and $Q_{\mathrm{total}}^{A}$ ($Q_{\mathrm{total}}^{B}$) are defined as the tail and total charge integral observed by light sensor A (B), respectively. The PSP was calculated only for the rod or bar that collected the largest charge in an event. The figure of merit (FOM) is then calculated as
%

%
\begin{equation} \label{eq_fom}
    \mathrm{FOM} \equiv \frac{\mu_{e} - \mu_{n}}{FWHM_{e}+FWHM_{n}},
\end{equation}
where $\mu_{e}$($\mu_{n}$) and $FWHM_{e}$($FWHM_{n}$) are the mean and the full-width-half-maximum values of the PSP distribution for electron (proton) recoils at any given electron equivalent energy.
$^{252}$Cf and $^{22}$Na data were used to optimize the PSP; 
we used pure electron-recoil data ($^{22}$Na) to characterize the lower PSP band before finding the upper PSP band in the mixed electron/proton-recoil data ($^{252}$Cf).
The optimized charge integration limits for the SiPM data were found to be [$t_{L}-40$~ns $\leq Q_{\mathrm{total}} \leq$ $t_{L}+1120$~ns] and [$t_{L}+76$~ns $\leq Q_{\mathrm{tail}} \leq$ $t_{L}+1120$~ns]. The optimized charge integration limits for the PMT data were found to be [$t_{L}$-20~ns~ $\leq Q_{\mathrm{total}} \leq$ ~$t_{L}$+1300~ns] and [$t_{L}$+28~ns~ $\leq Q_{\mathrm{tail}} \leq$ ~$t_{L}$+1300~ns]. 
The charge response differences from channel to channel caused by gain or optical coupling efficiency variations were corrected using a collimated $^{137}$Cs gamma-ray source directed at the center of the module.
Here, lead bricks were used for shaping the $^{137}$Cs source into a fan beam of $\sim$0.5-cm width. 
For each component of SANDD, each channel's charge response was equalized by scaling the position of its Compton continuum maximum to the channel with the largest Compton continuum maximum.

\begin{figure*}[ht]
\begin{multicols}{3}
    \includegraphics[width=1.0\linewidth]{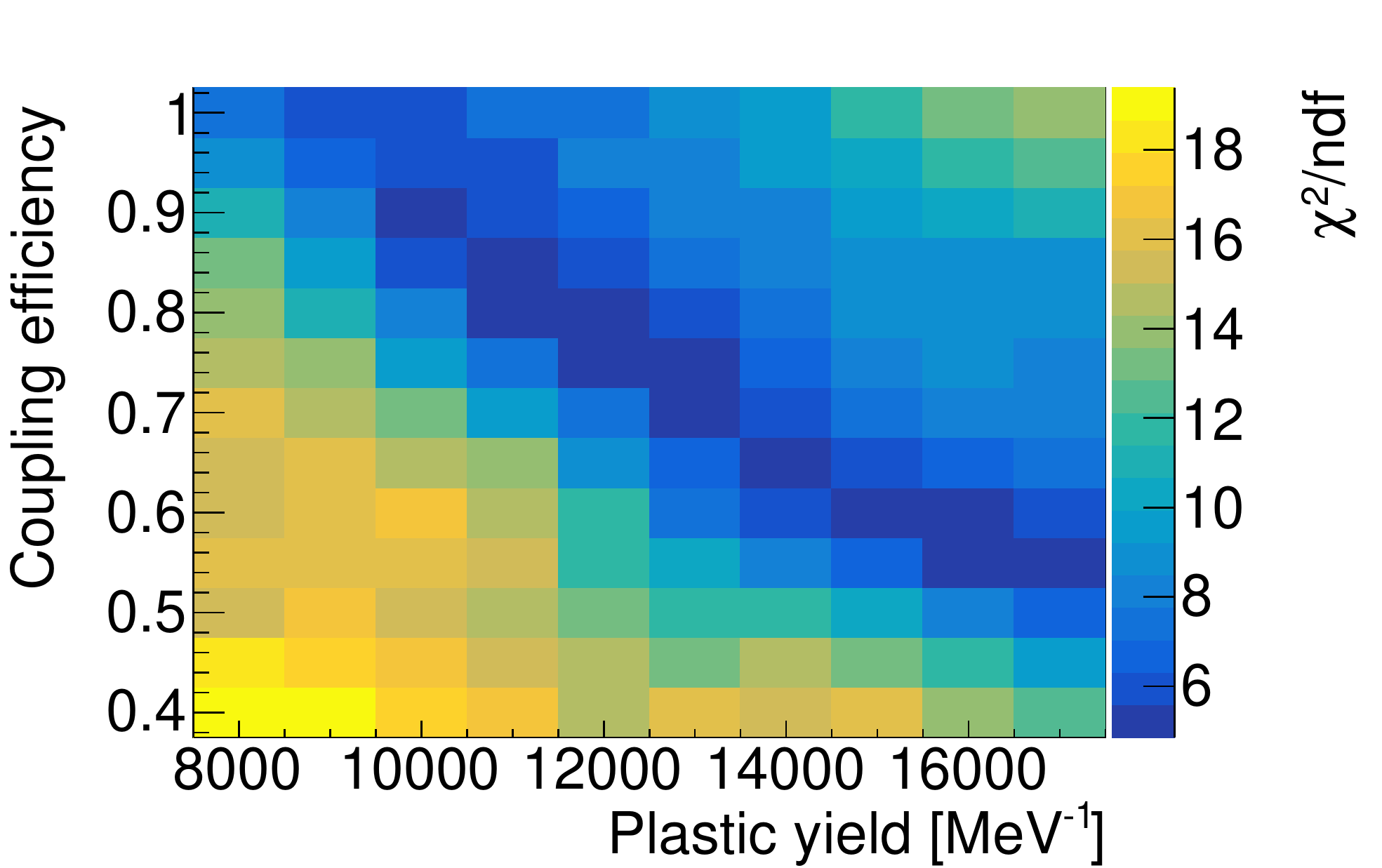} \par 
    \includegraphics[width=1.0\linewidth]{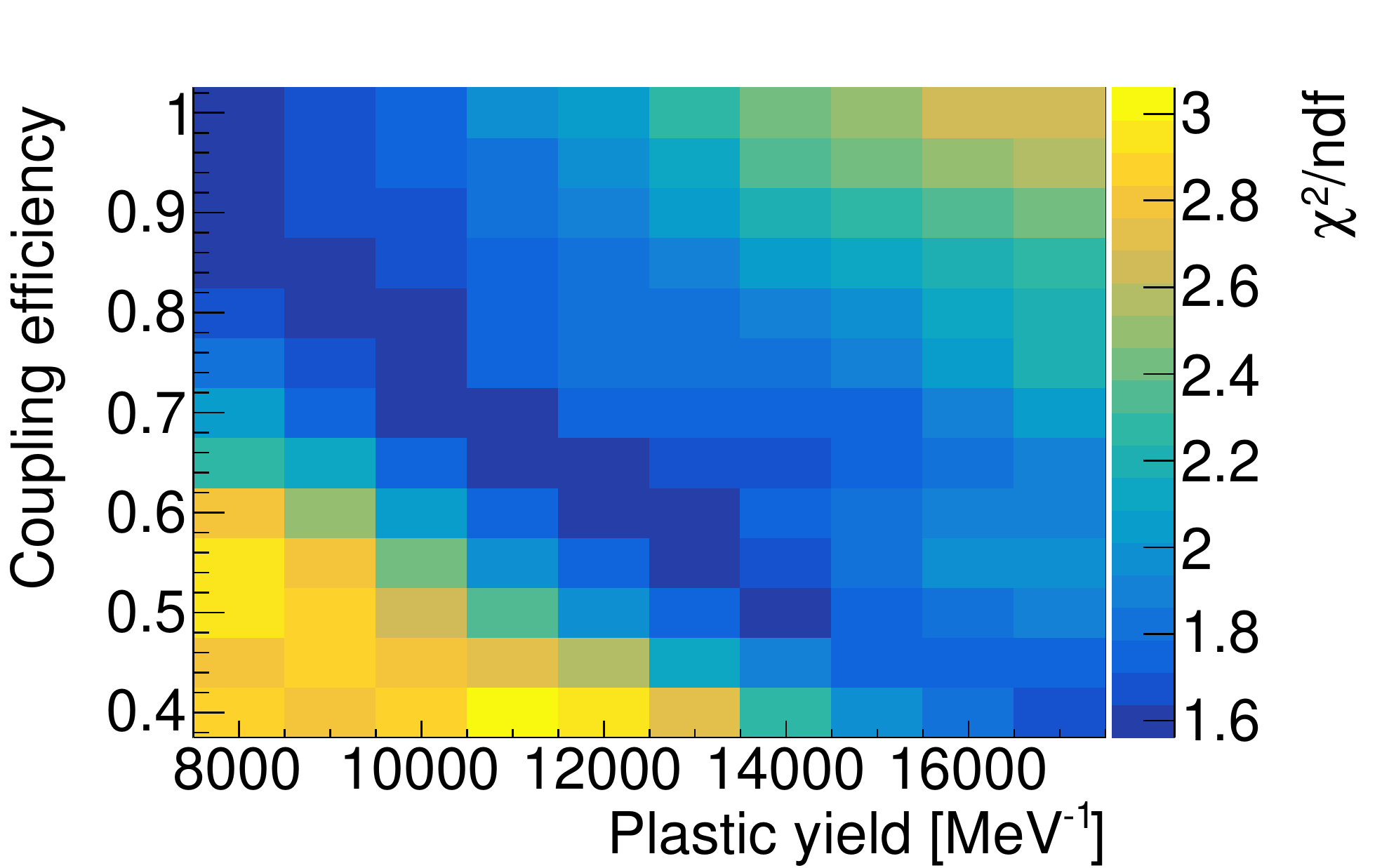} \par 
    \includegraphics[width=1.0\linewidth]{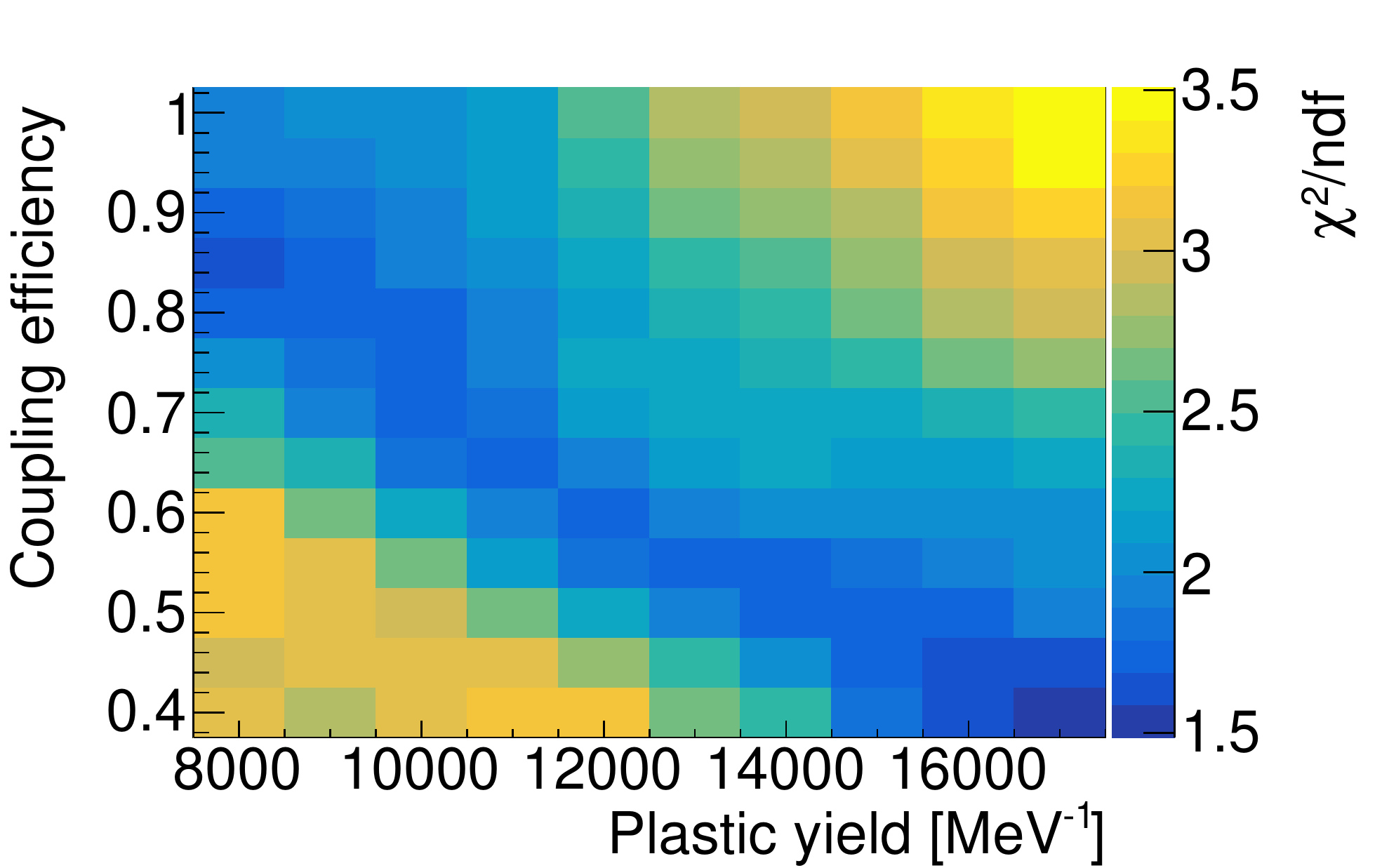} \par 
\end{multicols}
\begin{multicols}{3}
\begin{centering}
    (a) \par 
    (b) \par 
    (c) \par 
\end{centering}
\end{multicols}
\vspace{-0.3cm}
\caption{Distributions of goodness of fit ($\chi^2$/ndf) as a function of coupling efficiency and plastic yield for (a)~component 1, (b)~component 2, and (c)~component 3.}
\label{fig_chiLo}
\end{figure*}

\begin{figure*}[ht]
\begin{multicols}{3}
    \includegraphics[width=1.0\linewidth]{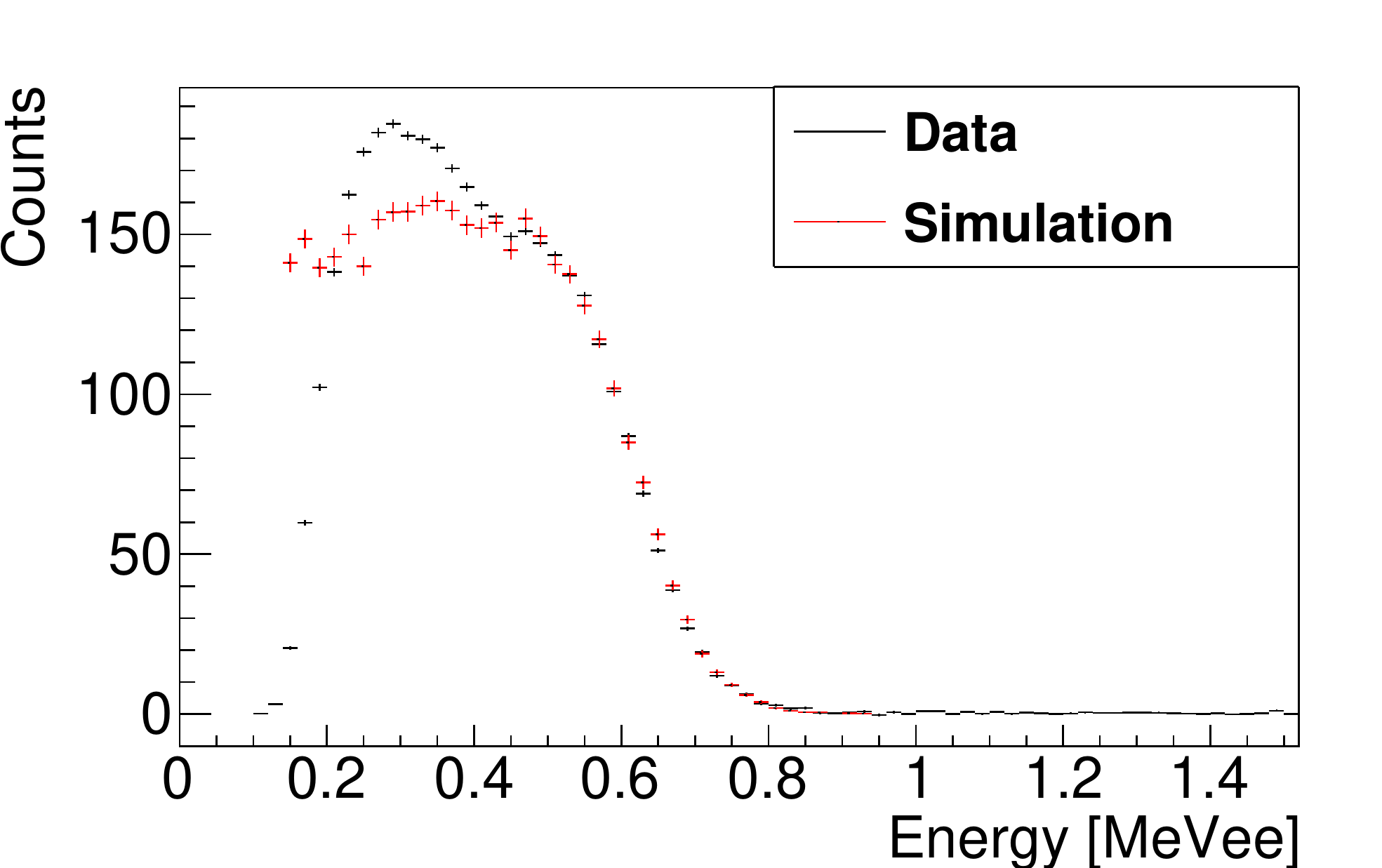} \par 
    \includegraphics[width=1.0\linewidth]{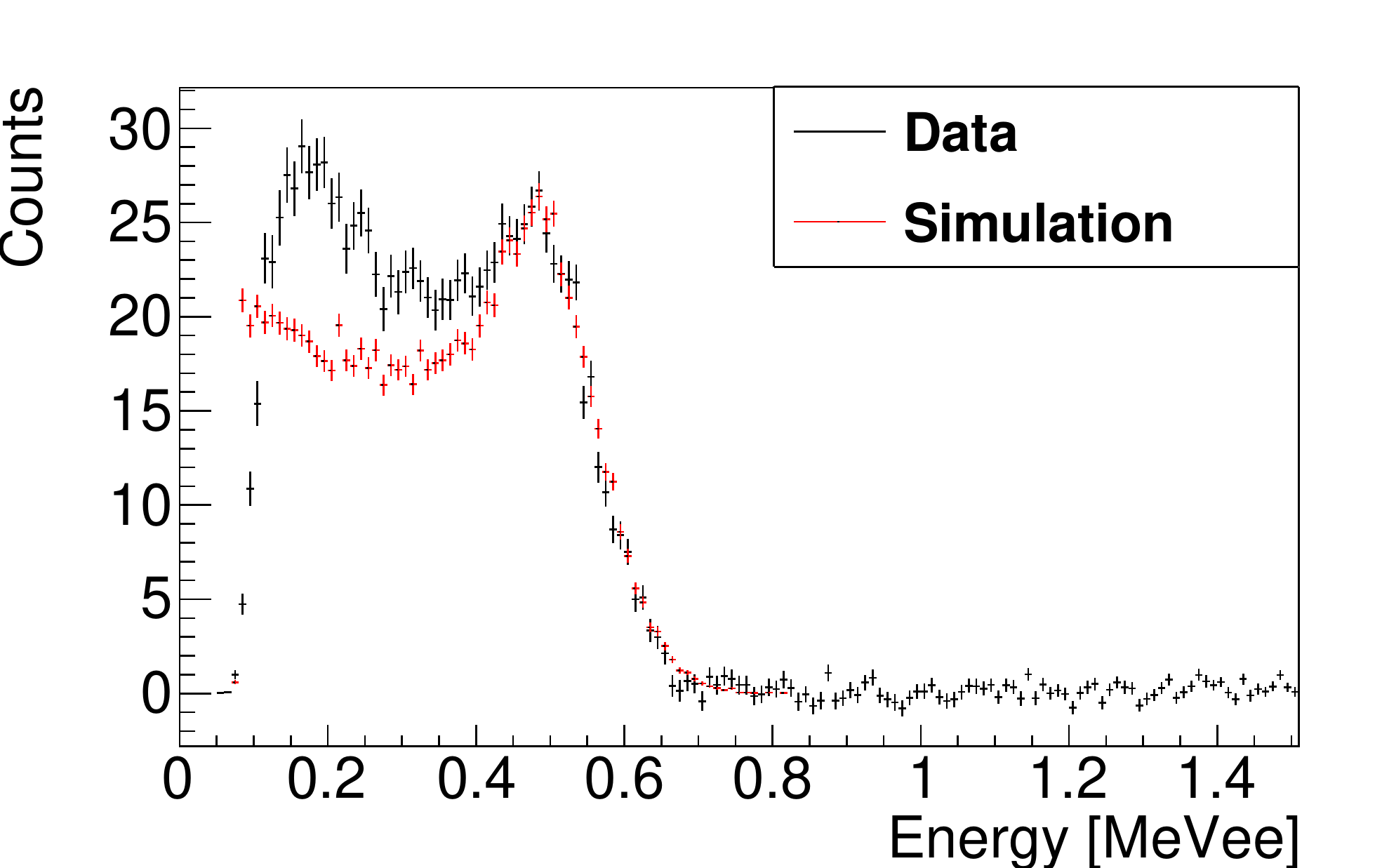} \par 
    \includegraphics[width=1.0\linewidth]{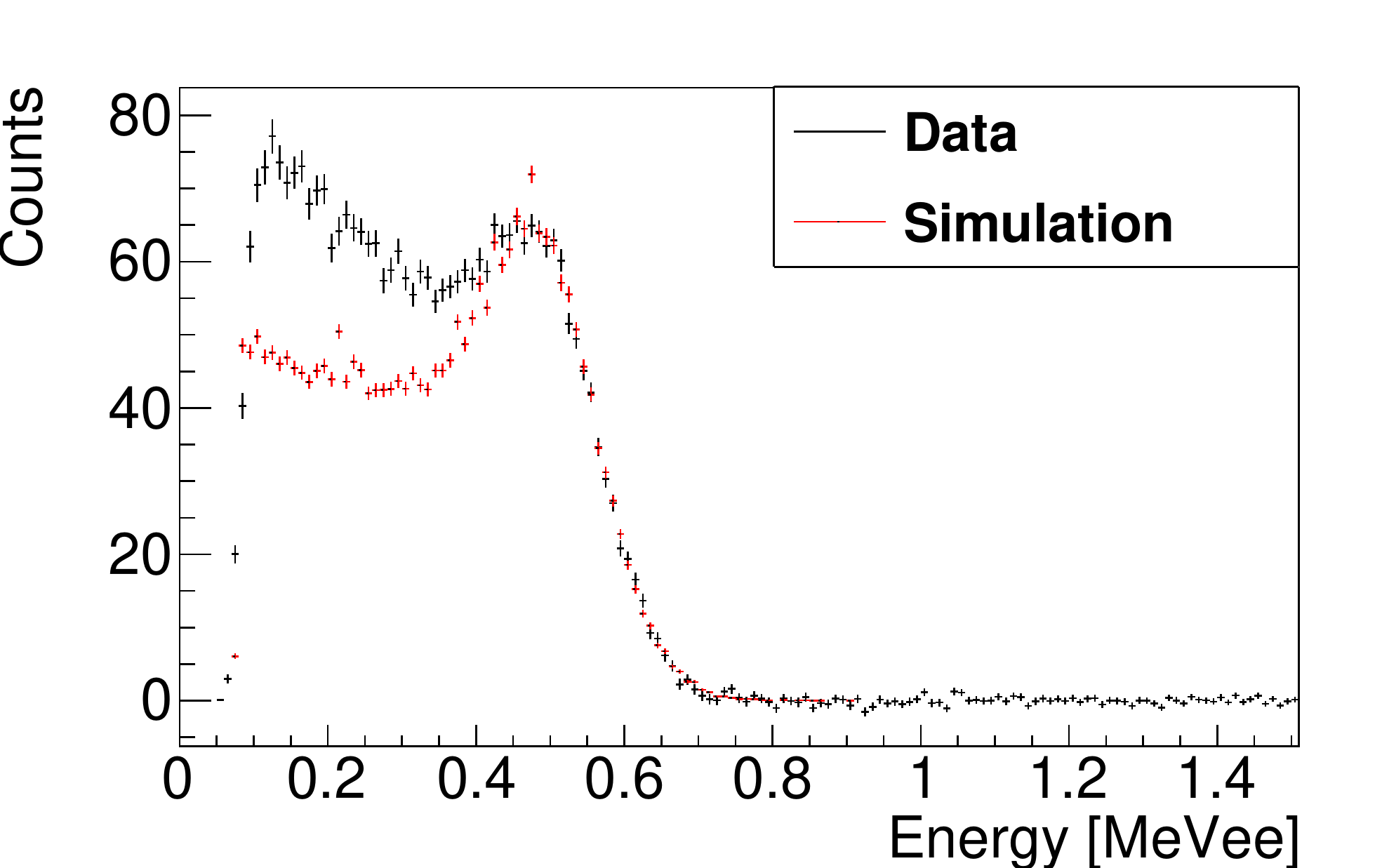} \par 
\end{multicols}
\begin{multicols}{3}
\begin{centering}
    (a) \par 
    (b) \par 
    (c) \par 
\end{centering}
\end{multicols}
\vspace{-0.3cm}
\caption{Example comparison between the simulated and the measured energy responses to a collimated $^{137}$Cs source for (a)~component 1, (b)~component 2, and (c)~component 3.}
\label{fig_lo}
\end{figure*}

To measure the apparent attenuation length of the scintillator, a series of collimated $^{137}$Cs measurements were taken at regular intervals along the length of each component. 
In each case, the location of Compton continuum maximum was identified by fitting the energy response with a Gaussian profile while varying the range of the fit to find the minimum $\chi^2$. 
The Compton continuum maximum position was identified as the mean of Gaussian profile that yielded the minimum $\chi^2$. To estimate the uncertainty, we varied the range of the fit until the $\chi^2$ value exceeds the 68\% confidence level (CL) of the minimum $\chi^2$; the uncertainty was the corresponding range of the mean of the Gaussian profile. 
Fig.~\ref{fig_abs} shows the charge collected at both light sensors as a function of source position for all three different types of component. Apparent attenuation lengths of 18$\pm$2~cm, 28$\pm$2~cm, and 31$\pm$2~cm were measured for component 1, component 2, and component 3, respectively.

In the Geant4 simulation, there are two parameters influence the apparent attenuation length: intrinsic attenuation length of the plastic and the surface roughness of the plastic. A simulation was performed to investigate the relative importance of these two characteristics. The surface roughness parameter was modeled with the 
sigma alpha 
parameter in Geant4~\cite{sigmaalpha}. In the large cross section bar simulations (components 2 and 3), scintillator bars were wrapped with the Teflon tape. The 
UNIFIED 
model was used with the Teflon reflection type set to lambertian distribution and scintillator reflection type set to specular lobe distribution. The 
sigma alpha 
parameter describes the degree of non specular reflection at surface in terms of a Gaussian distribution with a width of 
sigma alpha 
radians. 
The simulation placed a small air gap between the scintillator and the Teflon tape. The refractive index of the scintillator was assumed to be similar to polystyrene ($\sim$1.6 in the range of 480--680~nm). The shapes of the absorption and the emission spectra of the plastic were taken from measurements~\cite{Mabe_2019}.

\begin{figure}[ht]
\centering\includegraphics[width=1.0\linewidth]{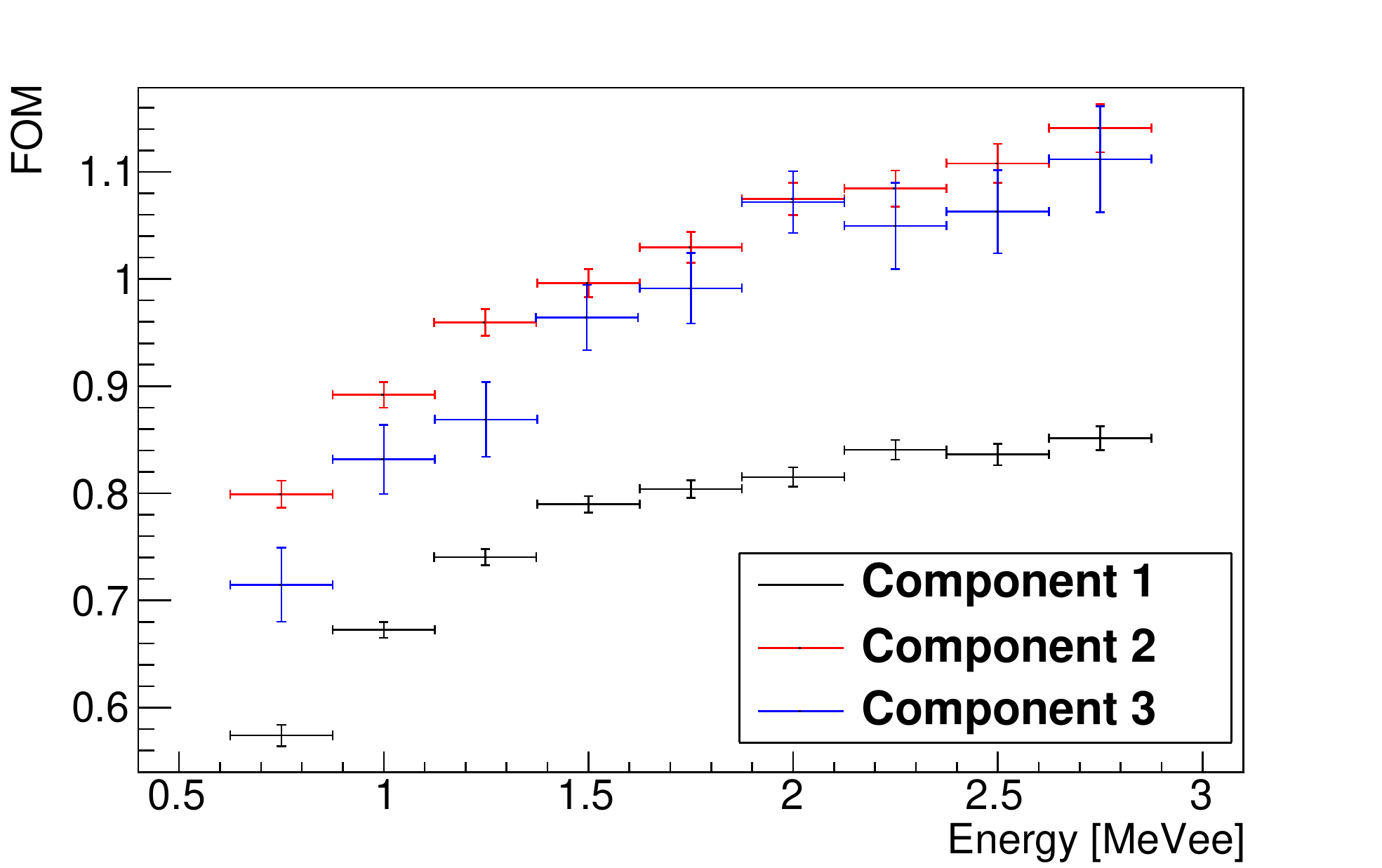} 
\caption{FOM as a function of energy. For the central module data (Component 1), the response was calculated based on the rod that collected the largest charge in an event.}
\label{fig_fom}
\end{figure}

\begin{figure*}
\begin{multicols}{3}
    \includegraphics[width=1.0\linewidth]{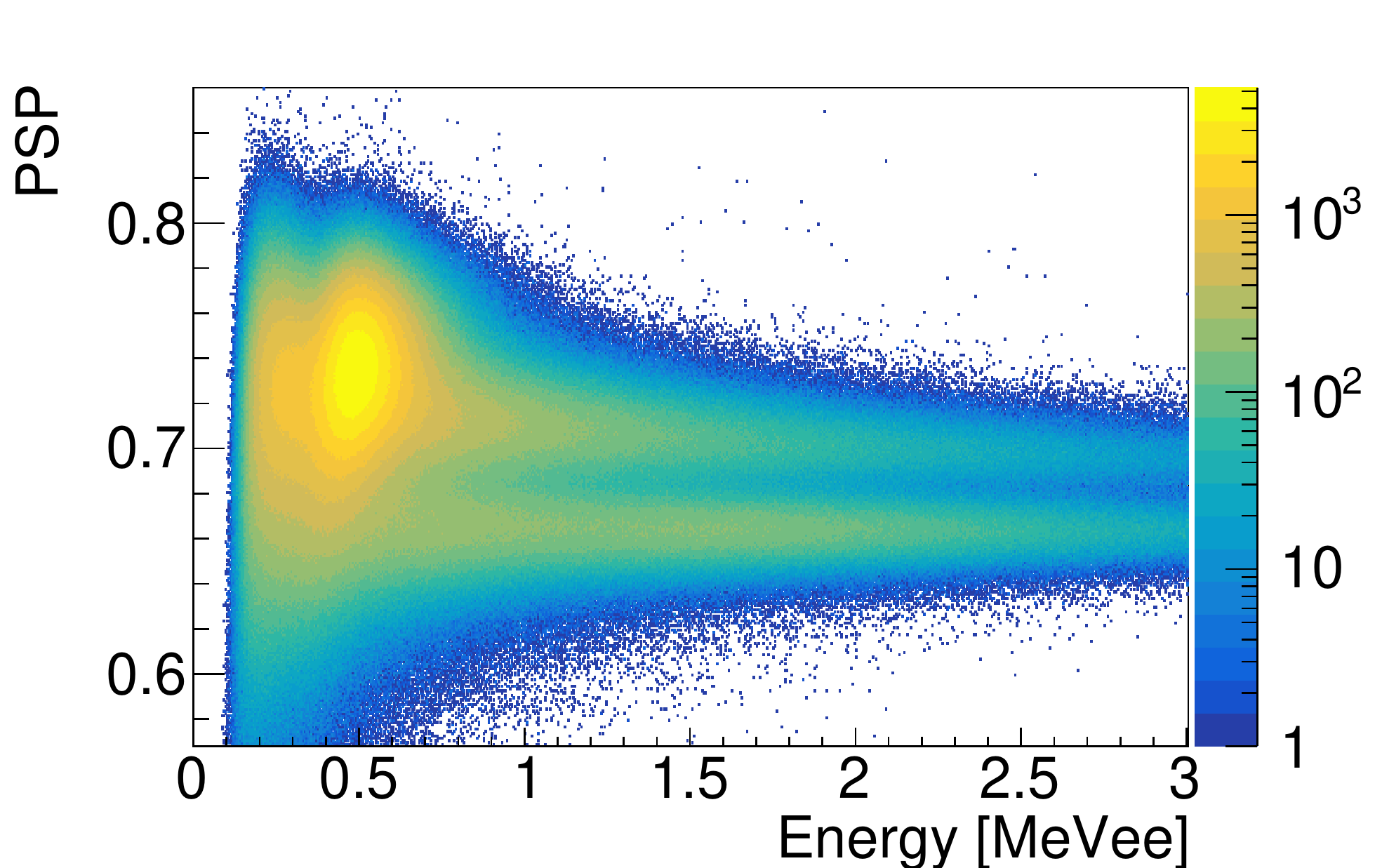} \par 
    \includegraphics[width=1.0\linewidth]{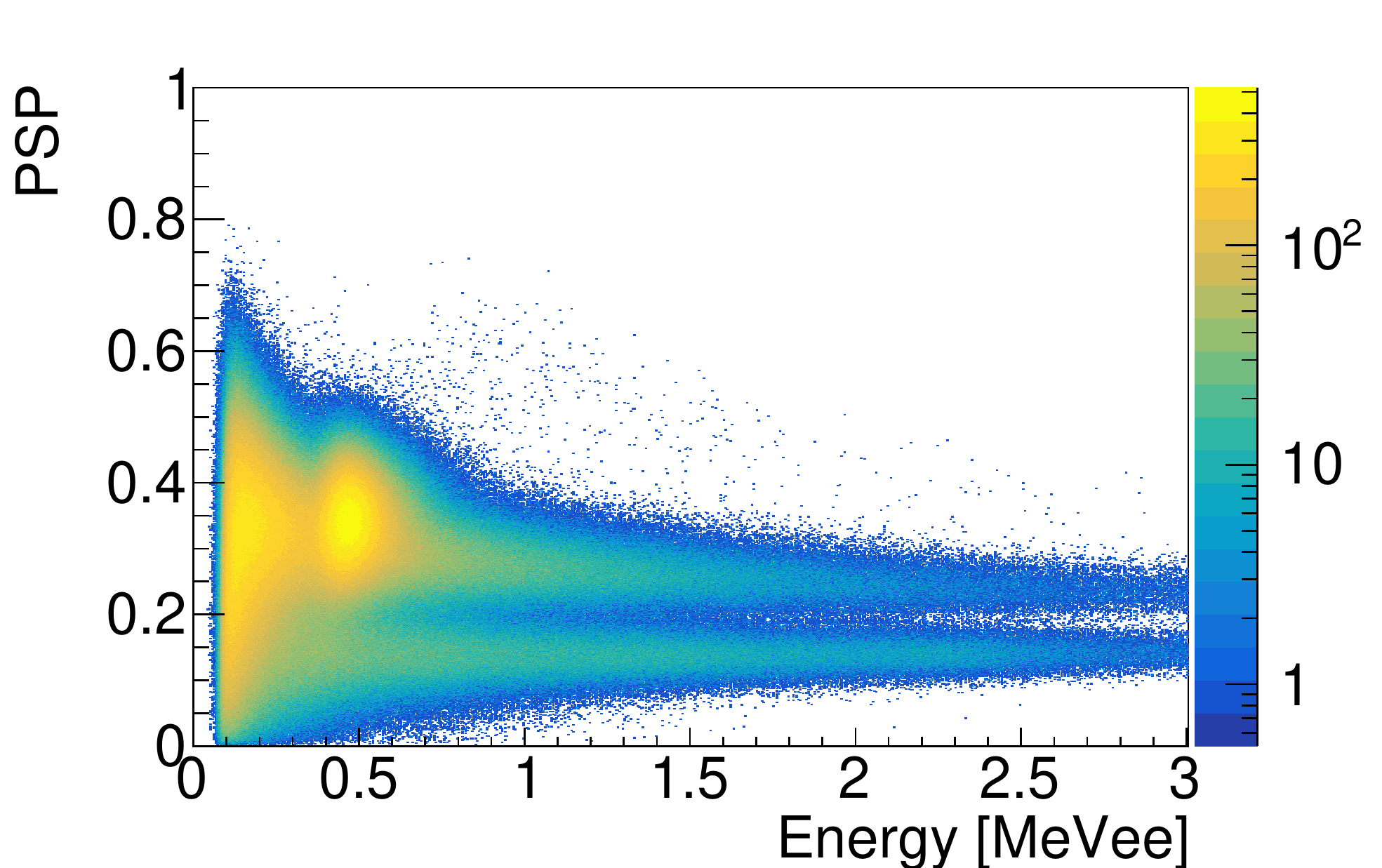} \par 
    \includegraphics[width=1.0\linewidth]{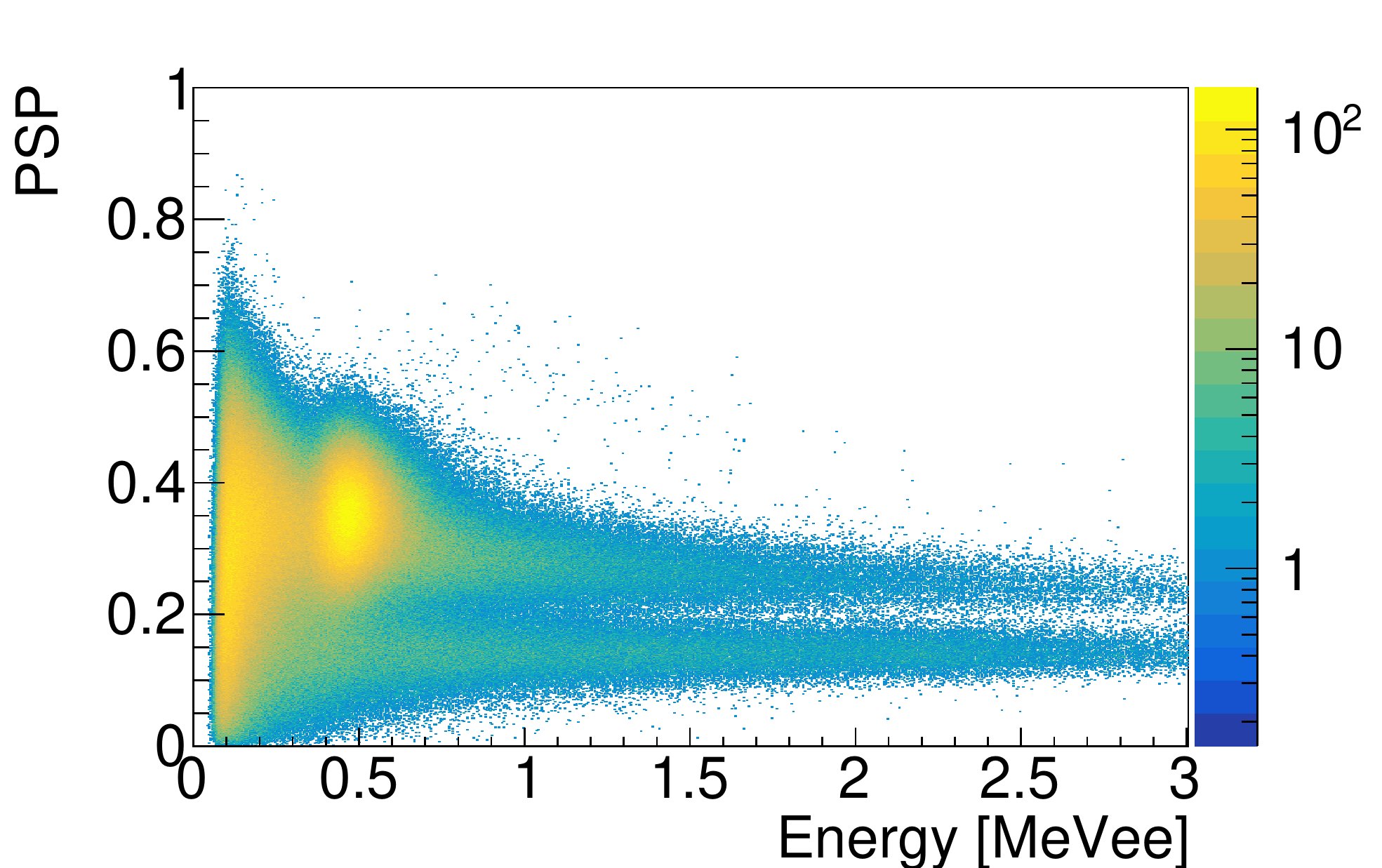} \par 
\end{multicols}
\begin{multicols}{3}
\begin{centering}
    (a) \par 
    (b) \par 
    (c) \par 
\end{centering}
\end{multicols}
\vspace{-0.3cm}
\caption{PSP-energy responses to $^{252}$Cf source of (a)~component 1, (b)~component 2, and (c)~component 3.}
\label{fig_psp}
\end{figure*}

The comparisons to simulation in Fig.~\ref{fig_abs} were obtained by simulating 477~keV electrons at the same positions as the data (with a 0.5-cm position spread). Note that the energy corresponding to the Compton continuum maximum of 662-keV $^{137}$Cs gamma rays is 477~keV~\cite{knoll}.
The energy depositions produce optical photons that are propagated through the scintillator material and arrive at the light sensors (SiPMs or PMTs). The Hamamatsu PMT quantum efficiency~\cite{hamamatsu_specSheet} and the SensL SiPM photon detection efficiency~\cite{sensl_specSheet} were then folded into the detector responses. The charge responses from the simulation and experiment were normalized to minimize the sum of the squares of the differences between them as shown in Fig.~\ref{fig_abs}.

\begin{figure}[ht]
\centering\includegraphics[width=1.0\linewidth]{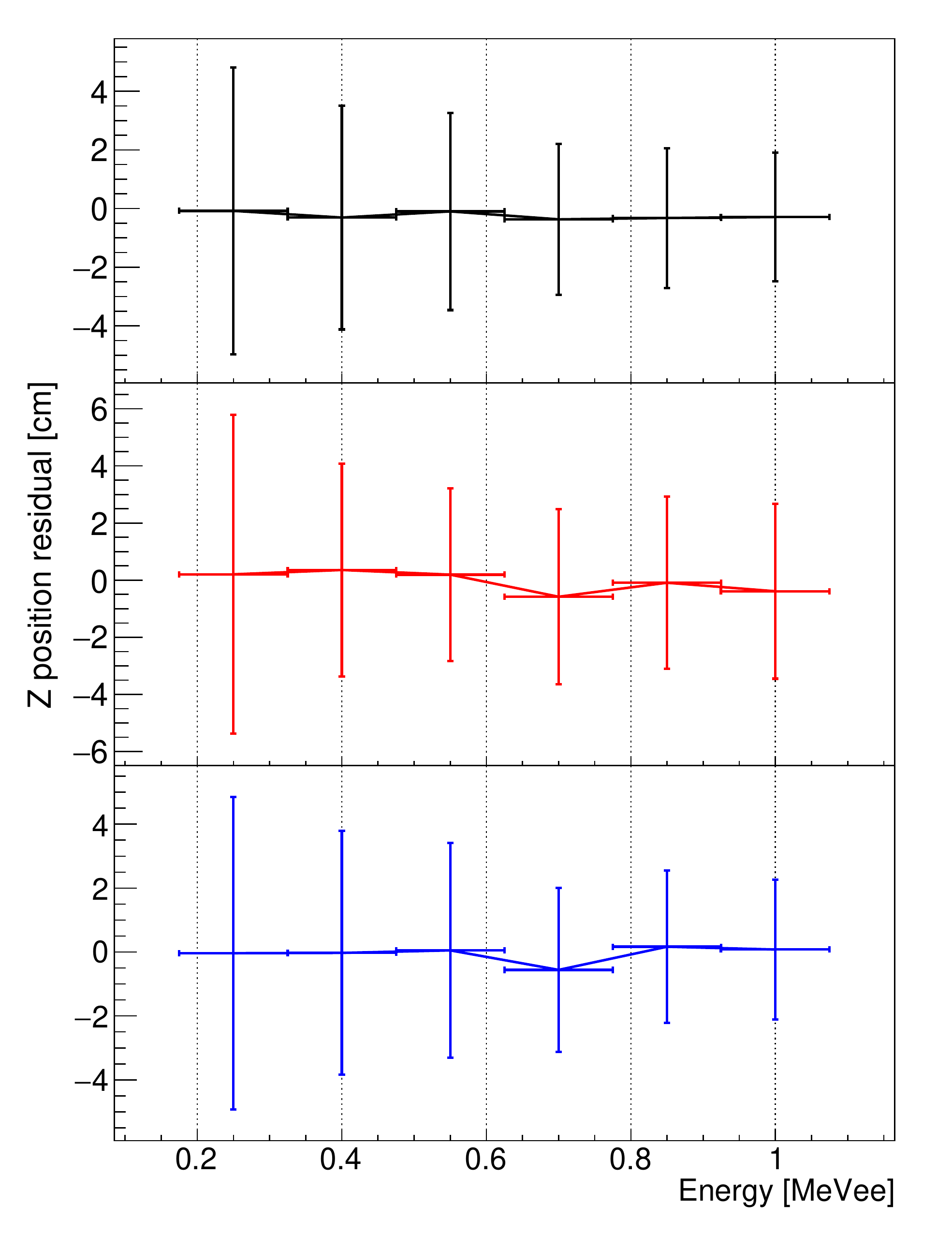}
\vspace{-0.3cm}
\caption{Z-position residual 
is plotted as a function of energy for component 1 (top), component 2 (middle), and component 3 (bottom). }
\label{fig:zPos}
\end{figure}

To determine the best combination of surface roughness and intrinsic attenuation length, independent $\chi^2$/ndf maps were obtained for light sensors A and B and for different components. Small differences in best-fit surface roughness and intrinsic attenuation length were then resolved using the Fisher method~\cite{fisher}, resulting in combined $\chi^2$/ndf map as shown in Fig.~\ref{fig_chiAbs}. 

\begin{figure*}[ht]
\begin{multicols}{3}
    \includegraphics[width=1.0\linewidth]{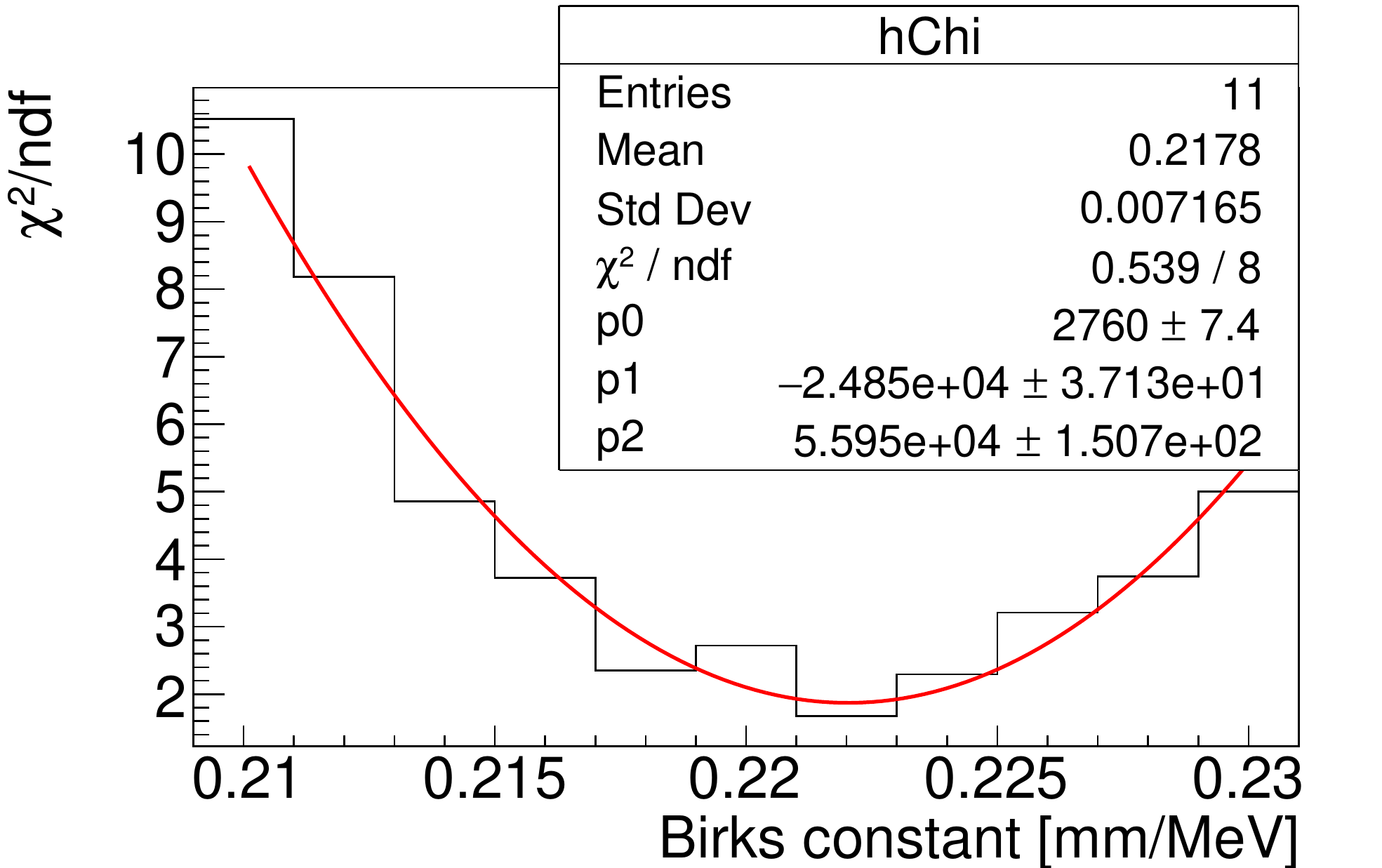} \par 
    \includegraphics[width=1.0\linewidth]{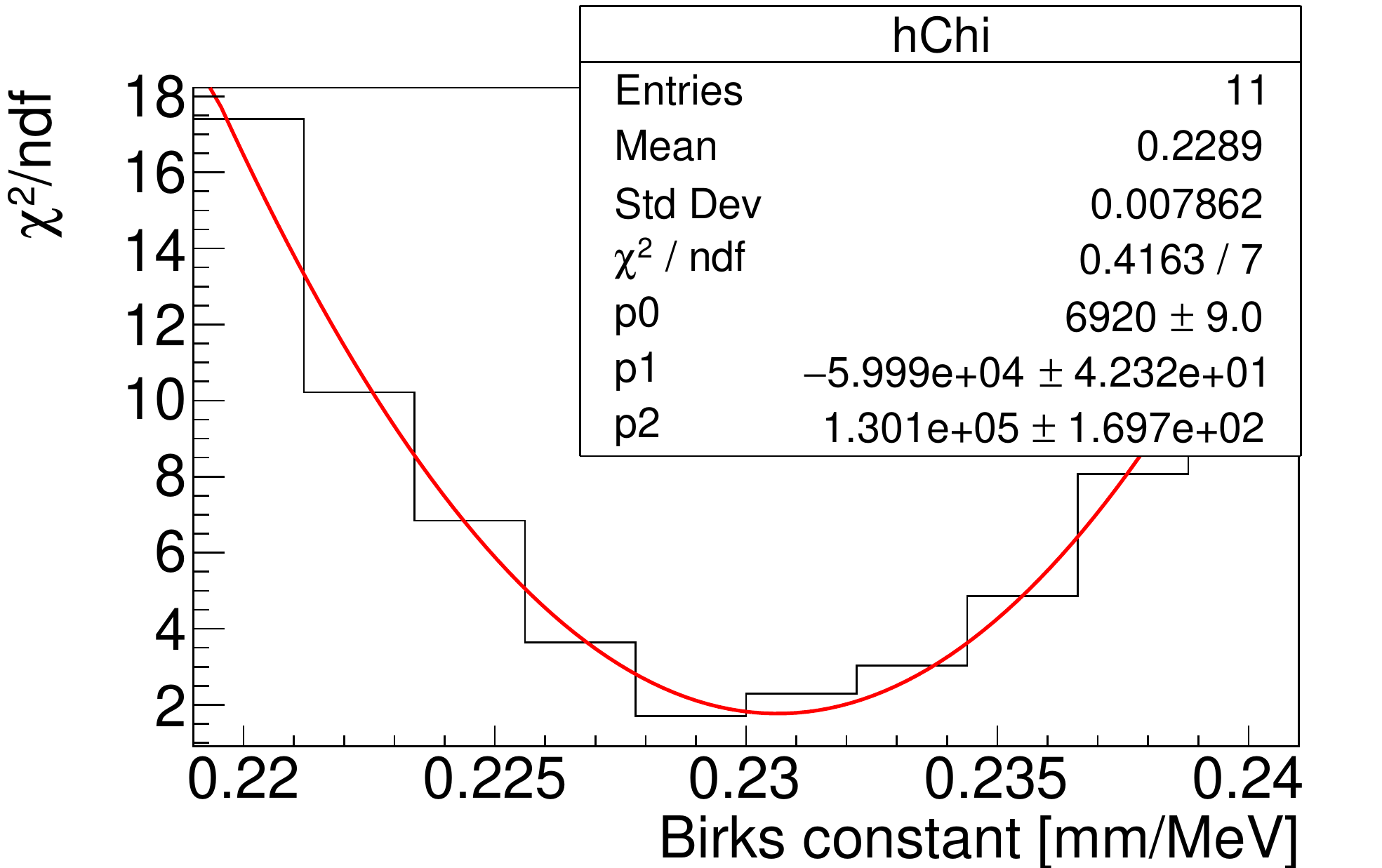} \par 
    \includegraphics[width=1.0\linewidth]{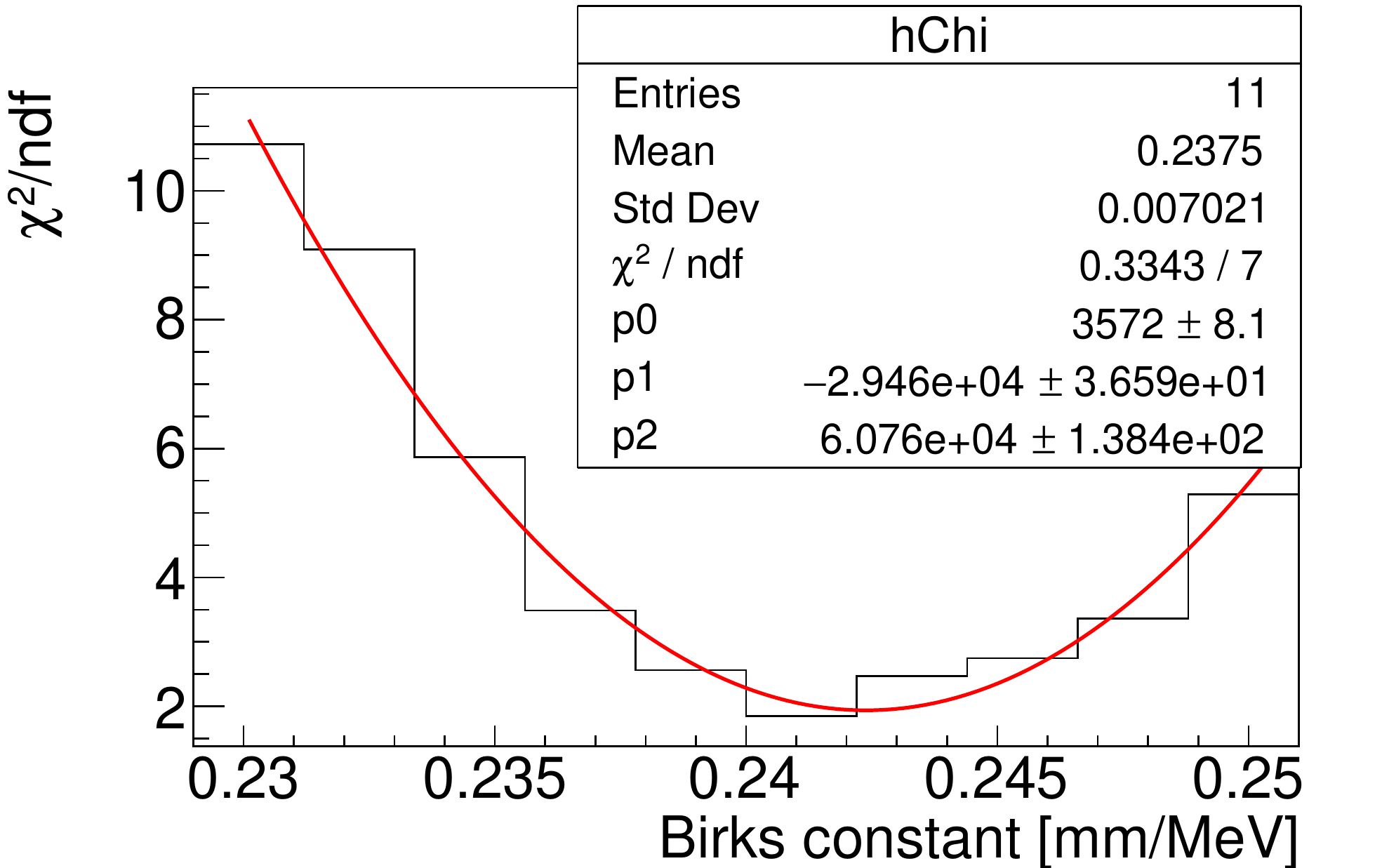} \par 
\end{multicols}
\begin{multicols}{3}
\begin{centering}
    (a) \par 
    (b) \par 
    (c) \par 
\end{centering}
\end{multicols}
\vspace{-0.3cm}
\caption{Goodness of fit ($\chi^2$/ndf) as a function of Birks constant for component 1 (a), component 2 (b), and component 3 (c). The parameters of second--order polynomial fit ($p_0$ + $p_1 x$ + $p_2 x^2$) are shown.}
\label{fig_chiCap}
\end{figure*}

\begin{figure*}[ht]
\begin{multicols}{3}
    \includegraphics[width=1.0\linewidth]{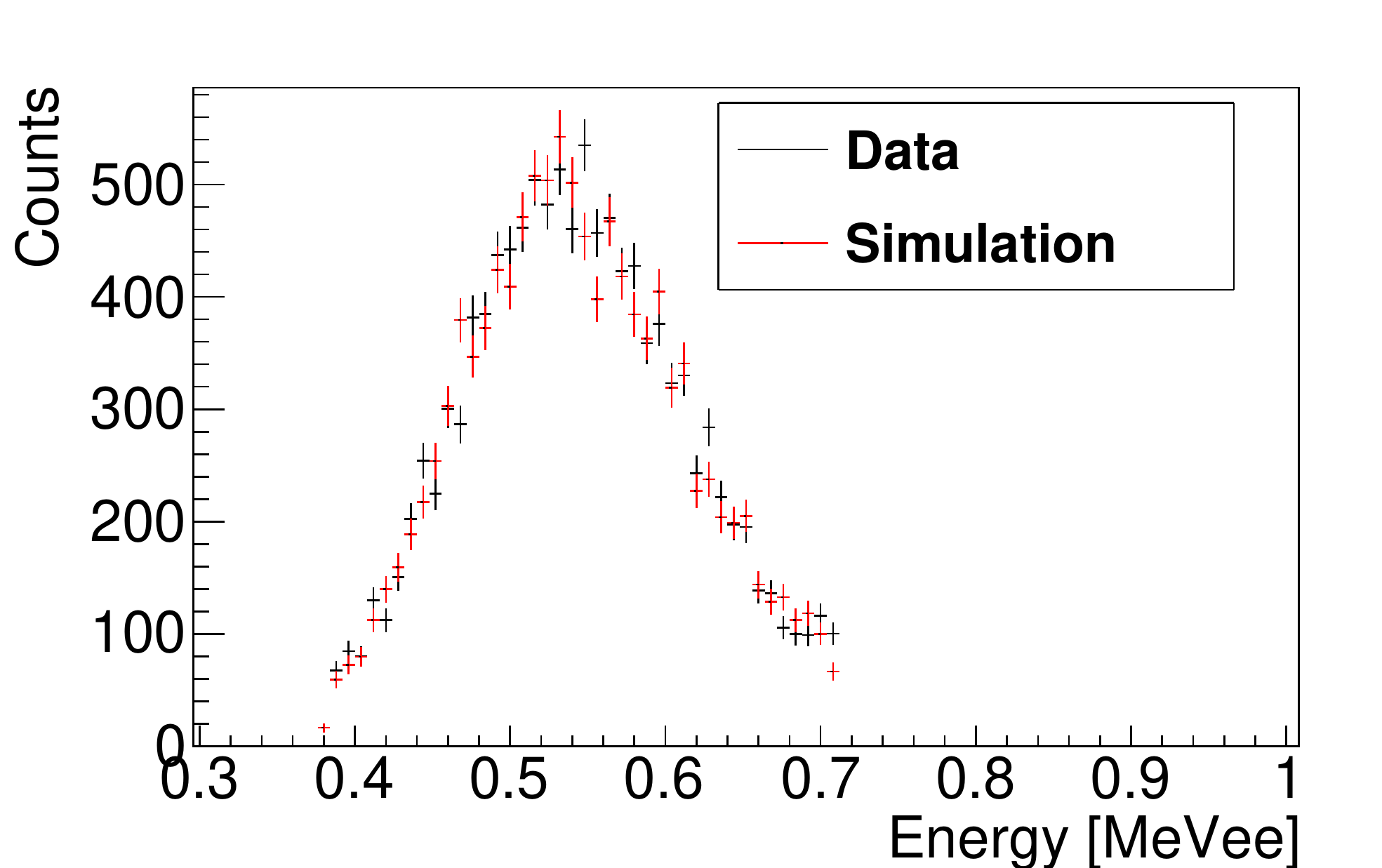} \par 
    \includegraphics[width=1.0\linewidth]{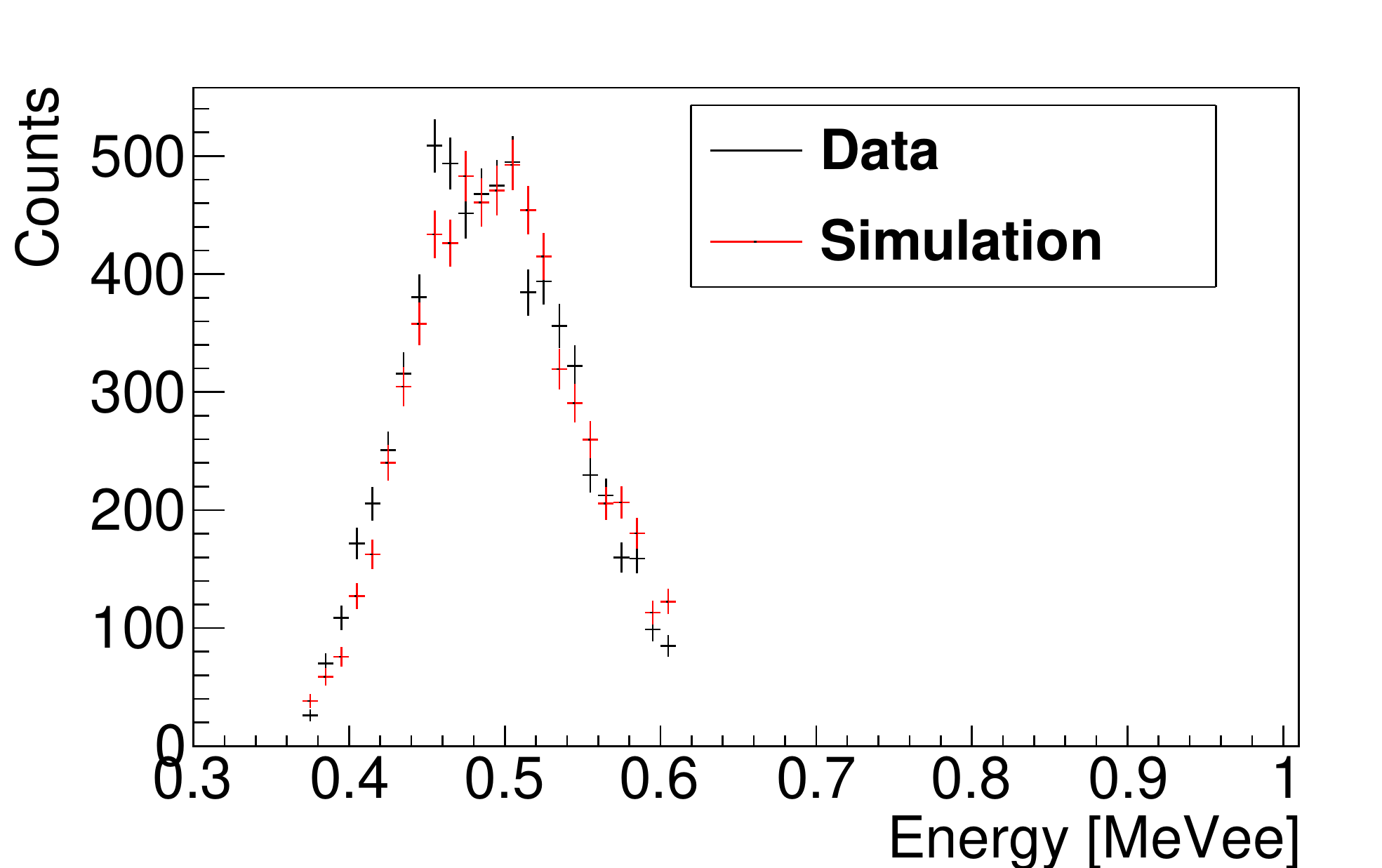} \par 
    \includegraphics[width=1.0\linewidth]{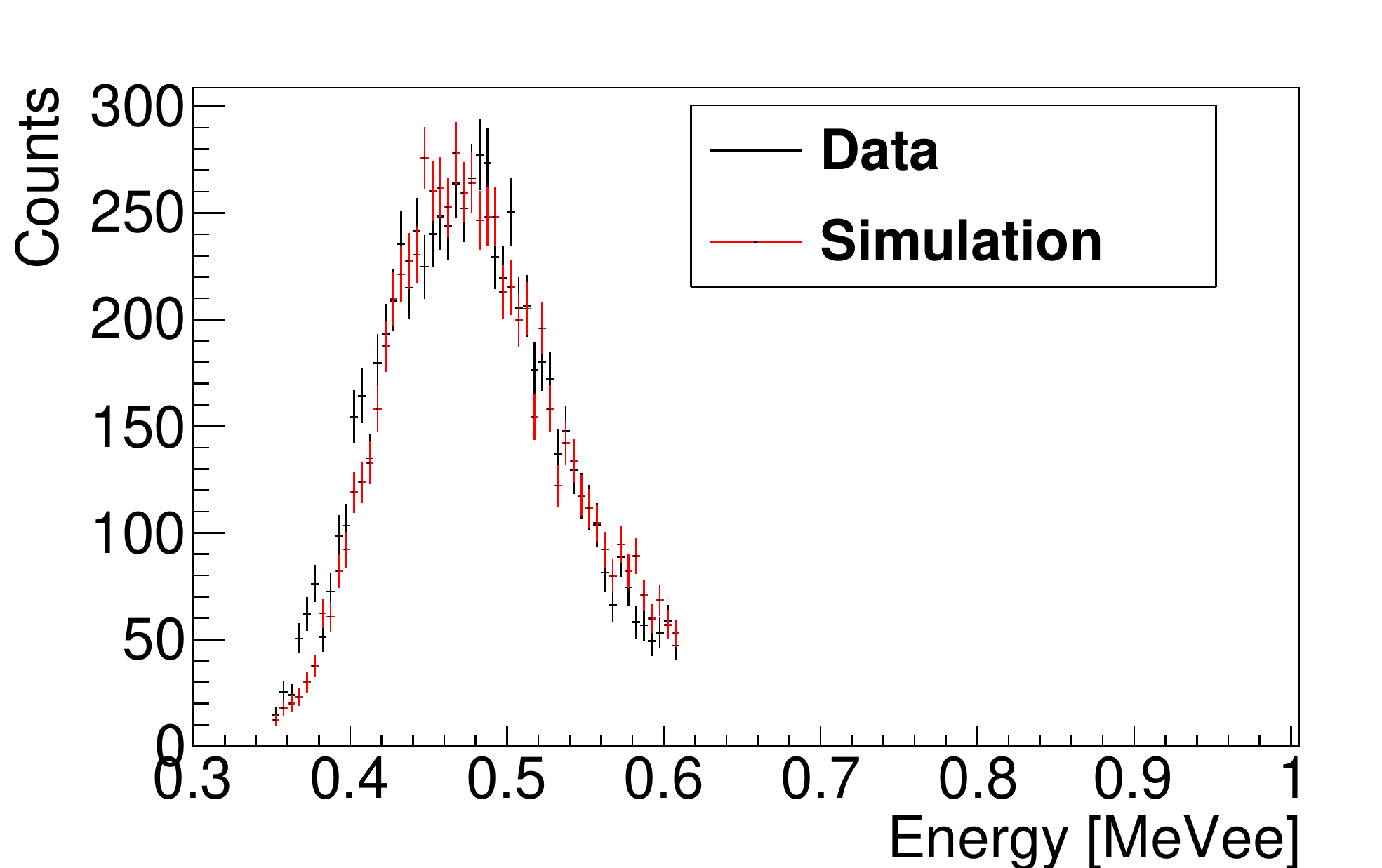} \par 
\end{multicols}
\begin{multicols}{3}
\begin{centering}
    (a) \par 
    (b) \par 
    (c) \par 
\end{centering}
\end{multicols}
\vspace{-0.3cm}
\caption{Comparison between the simulated and the measured energy responses to $^{252}$Cf neutrons for (a)~component 1, (b)~component 2, and (c)~component 3. The range of fit was set at 2~sigma from the mean energy response to neutron capture.}
\label{fig_cap}
\end{figure*}

To determine the range of acceptable surface roughness and intrinsic attenuation length, a series of 1D scans were performed over the range of  
sigma alpha 
values. 
Each scan yielded a $\chi^2$ plot over a range of intrinsic-attenuation-length. 
Each plot was then fitted with a second-degree parabolic function to determine the intrinsic attenuation length parameter corresponding to the $\chi^2$ minimum at that surface roughness. 
For each such minimum, the range of acceptable intrinsic attenuation length parameters were the ones that yielded a $\chi^2$ value within 68\% CL of the minimum value.
This generated a range of acceptable $\chi^2$ values (within 68\% CL) over a range of attenuation length and sigma alpha values. 
The minima of all the parabolic fits were taken as the best fit values and the uncertainty was given by the range of acceptable values.
The chosen 
sigma alpha 
and intrinsic attenuation length parameters are 0.12$\pm$0.02 and 72.6$\pm$10.3~cm, respectively (Table.~\ref{tab_param}).

The simulation framework was also tuned
in terms of light yield of the scintillator and the coupling efficiency between the light sensor and the scintillator by comparing the measured and simulated energy responses to a center collimated $^{137}$Cs source. To eliminate the dependence on the event position, the energy was calculated as follows:
\begin{equation} \label{Eq:EaEb}
    E = \sqrt{E_A E_B},
\end{equation}
where $E_A$ is the charge collected at light sensor A, $E_B$ is the charge collected at the opposite light sensor B, and $E$ is proportional to the energy deposited in the target volume. 
Equation.(\ref{Eq:EaEb}) assumes that light transport within each rod or bar behaves exponentially.
Figure~\ref{fig_lo} shows the measured background-subtracted energy responses to a collimated $^{137}$Cs source. The energy resolution values of approximately 22\%, 16\%, and 18\% were observed at the Compton continuum maximum of 662-keV gamma rays for component 1, component 2, and component 3, respectively. 
The measured energy responses were then compared to the simulated ones within [0.4,1.0]~MeVee energy range to obtain $\chi^2$ maps as shown in Fig.~\ref{fig_chiLo}.
A similar optimization was performed to tune the scintillator light yield and scintillator/light sensor coupling efficiencies.
The two parameters are fully correlated, and hence a strong correlation as shown in the band of good fits extending from top left to down right can be observed (Fig.~\ref{fig_chiLo}). Since we cannot estimate the two parameters independently, we report the combined value of both parameters as shown in Table~\ref{tab_param}.
An example comparison between the measured and simulated energy responses for one of the best fits is shown in Fig.~\ref{fig_lo}.
The discrepancy below 0.4~MeVee may be due to the low-energy scattered gamma-ray interactions in the target volume that are not captured by the simulation.
For example, the 662-keV gamma rays may scatter off the detector's surrounding materials and deposit their energy in the target volume.

Figure~\ref{fig_psp} shows the measured central module and the large cross section bars PSP-energy response to an uncollimated $^{252}$Cf source. 6"-thick lead were placed between the detector and the $^{252}$Cf source to reduce the gamma-ray flux. 
In each plot, two PSP bands are observed across the whole energy range. The upper and lower PSP bands correspond to proton and electron recoils, respectively. A $^6$Li neutron capture island is observed in each case at approximately 0.47~MeVee. 
The measured neutron island shape is shown in Fig.~\ref{fig_cap}. Gaussian profile was used to estimate the widths of the neutron island and the widths were found to be 0.08~MeV, 0.06~MeV, and 0.06~MeV for components 1, 2, and 3, respectively. The Gaussian shape of the capture island with respect to energy suggests that most captures result in both the alpha and triton fully depositing their energies within the same detector segment.

We also tuned the Birks parameter for the plastic scintillator in the simulation over a range of possible Birks parameters by comparing the measured response to $^{252}$Cf neutron captures with the simulated response as shown in Fig.~\ref{fig_chiCap}.
In doing so, 
sigma alpha
, intrinsic attenuation length parameter, coupling efficiency, and light yield were fixed to values shown in Table~\ref{tab_param}. The aforementioned method to determine the optimal simulation parameters and their associated uncertainties yielded the accepted ranges of the Birks constant, as shown in Table~\ref{tab_param}.
Comparison between the measured and the simulated neutron island for the best fit is shown in Fig.\ref{fig_cap}.

\begin{figure*}[h]
\begin{multicols}{3}
\noindent
    \includegraphics[width=\linewidth]{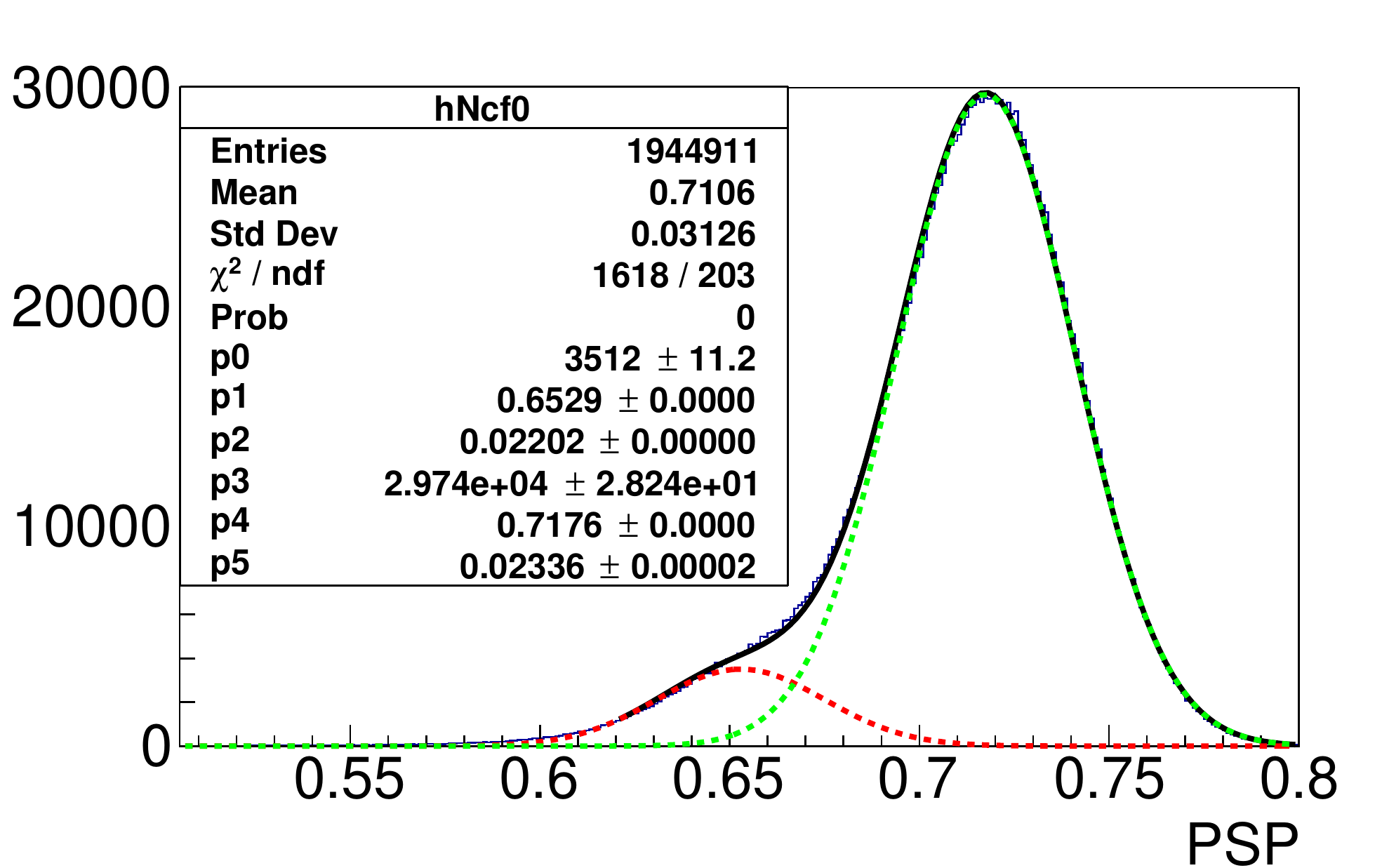} \par 
    \includegraphics[width=\linewidth]{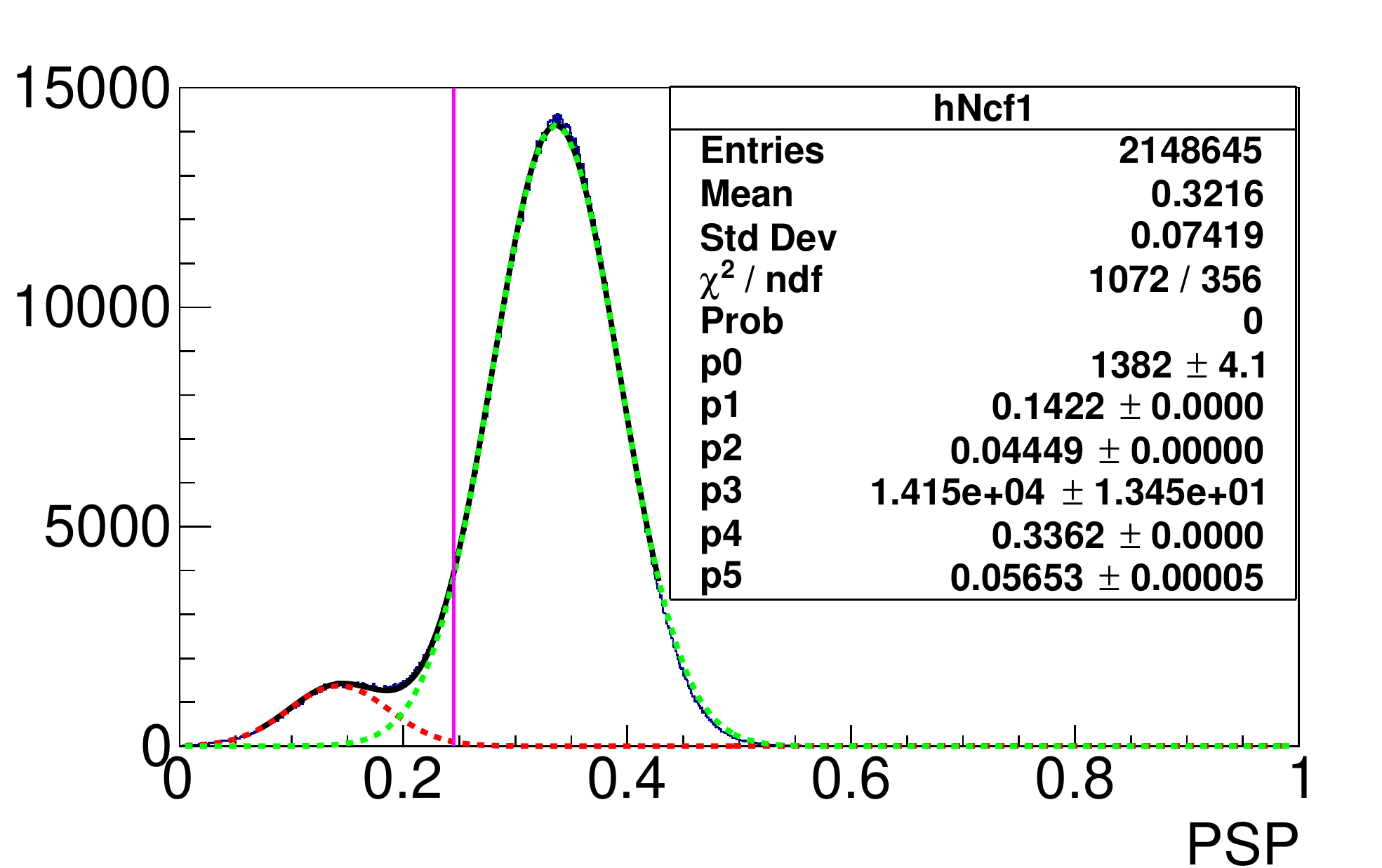} \par 
    \includegraphics[width=\linewidth]{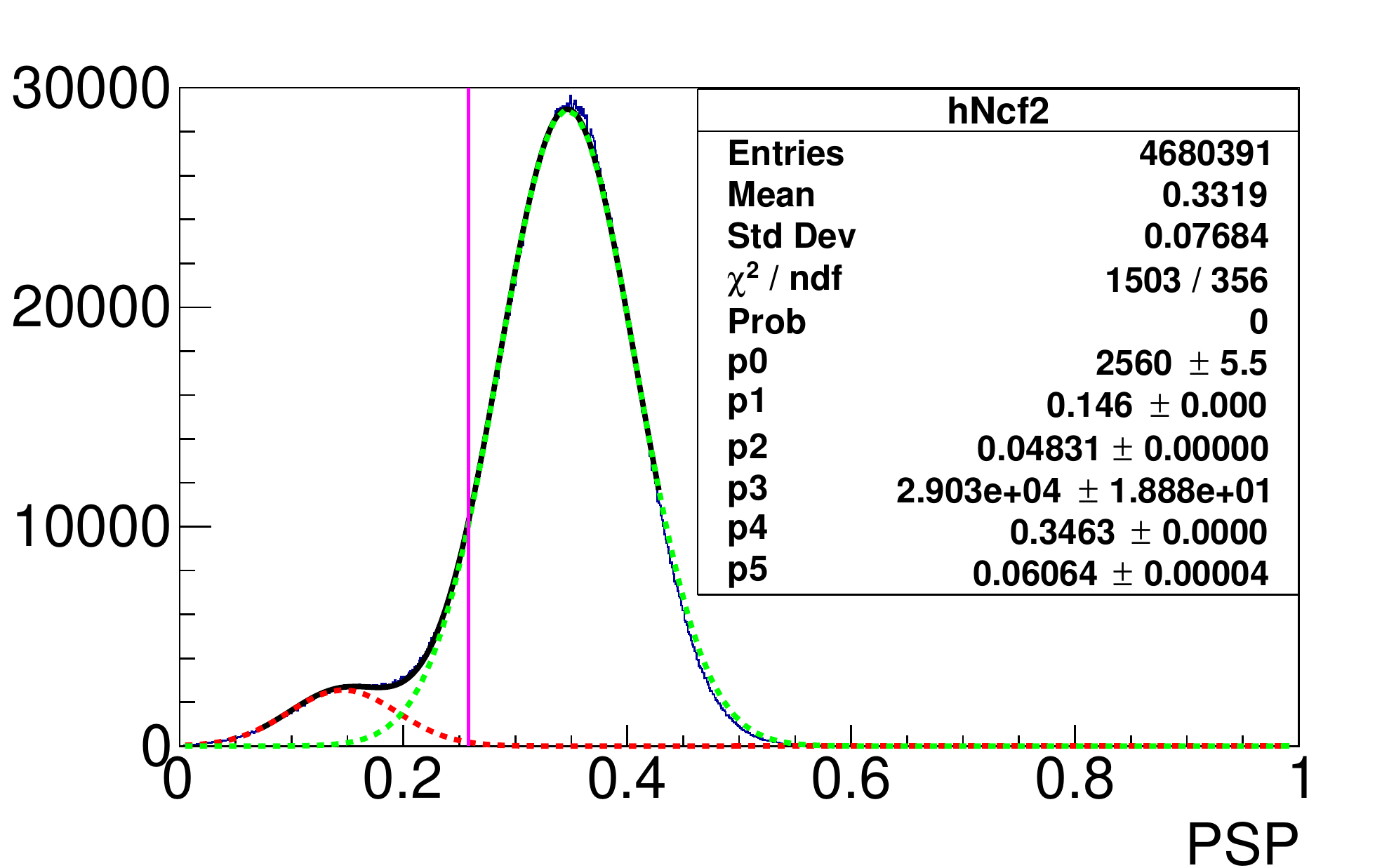} \par 
\end{multicols}
\begin{multicols}{3}
\noindent
\begin{centering}
    (a) \par 
    (b) \par 
    (c) \par 
\end{centering}
\end{multicols}
\vspace{-0.3cm}
\caption{PSP distribution for component 1 (a), component 2 (b), and component 3 (c) with energy range set at 2~sigma from the mean energy response to neutron capture. The PSP cuts to select neutron events in component 2 and 3 are shown (solid vertical line). The PSP cut to select neutron events in component 1 is not shown since the cut was determined separately for each rod. The dashed red and green lines correspond to electron recoils and thermal neutron captures in SANDD, respectively.}
\label{fig_pspNeu}
\end{figure*}

\begin{figure*}[ht]
\begin{multicols}{3}
\noindent
    \includegraphics[width=\linewidth]{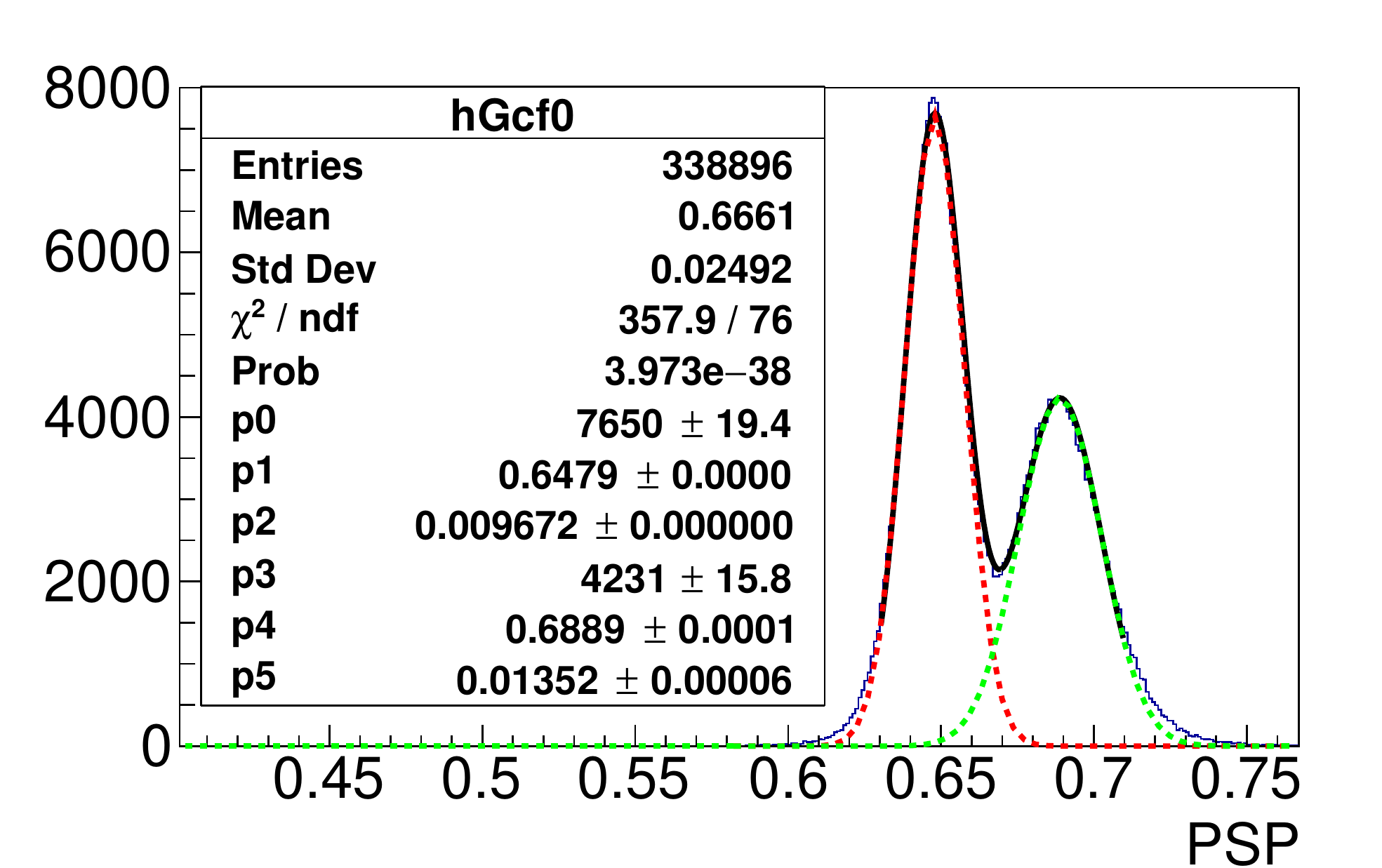} \par 
    \includegraphics[width=\linewidth]{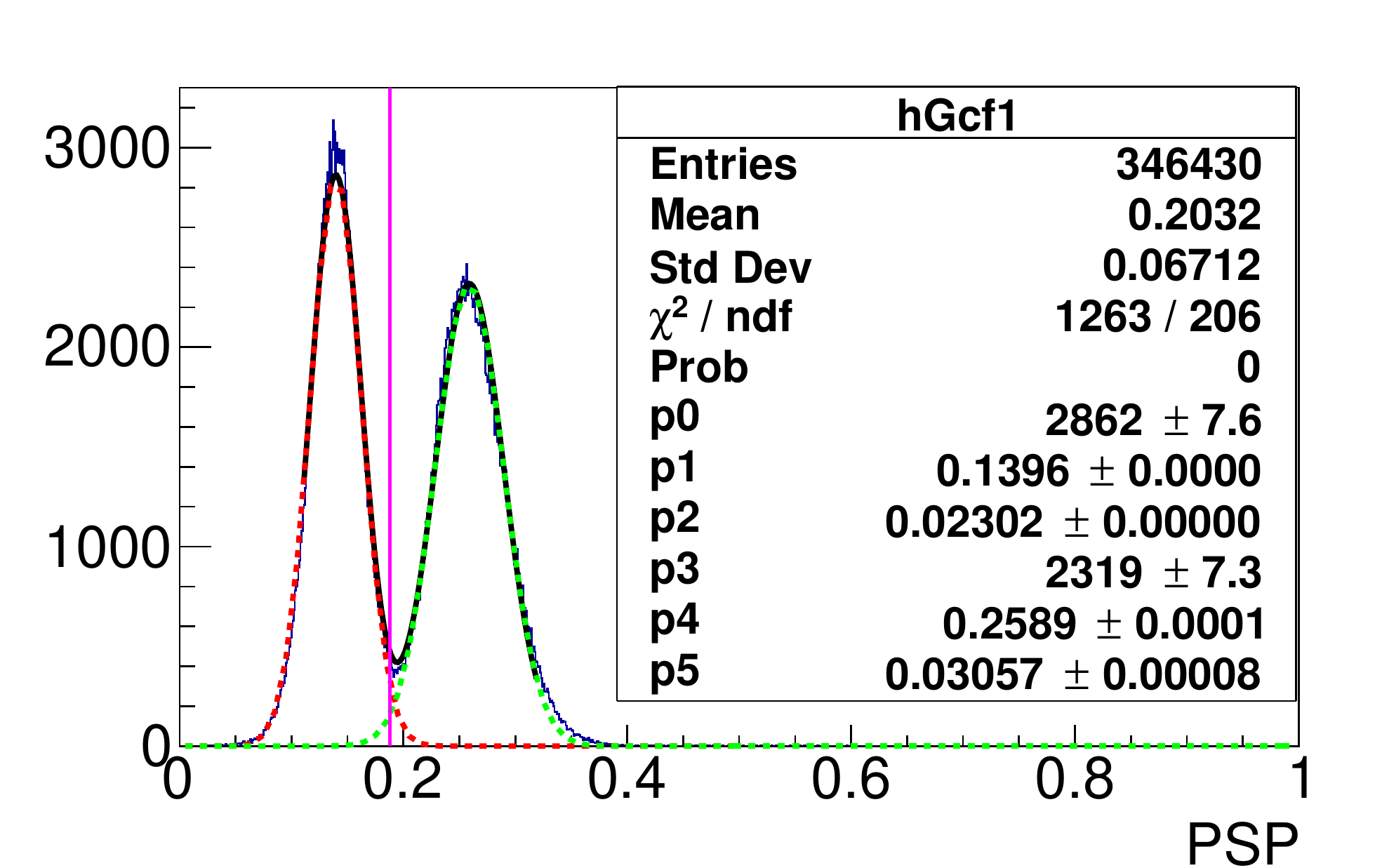} \par 
    \includegraphics[width=\linewidth]{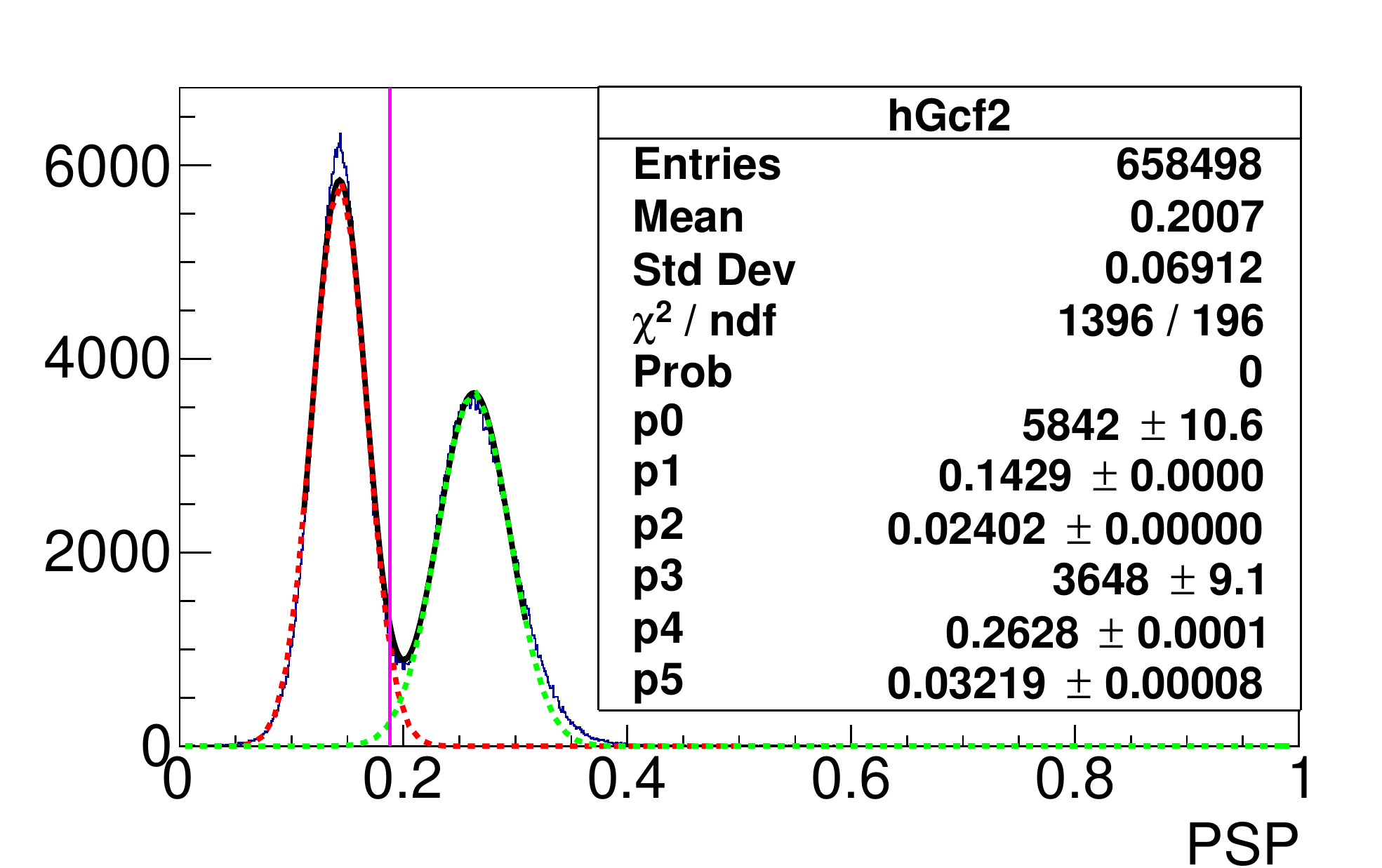} \par 
\end{multicols}
\begin{multicols}{3}
\noindent
\begin{centering}
    (a) \par 
    (b) \par 
    (c) \par 
\end{centering}
\end{multicols}
\vspace{-0.3cm}
\caption{PSP distribution with 1 -- 7~MeV energy range for component 1 (a), component 2 (b), and component 3 (c).  The PSP cuts used to select gamma events in component 2 and 3 are shown (solid vertical line). The PSP cut to select gamma events in component 1 is not shown since the cut was determined separately for each rod. The dashed red and green lines correspond to electron and proton recoils in SANDD, respectively. }
\label{fig_pspGam}
\end{figure*}

The FOM, as defined in Eq.~(\ref{eq_fom}), was used to quantify
the relative separation of the PSP for neutron and gamma rays. The FOM is presented as a function of energy in Fig.~\ref{fig_fom} for each of the  components. Components 1, 2, and 3 have FOM values of 0.67, 0.89, and 0.83 at 1~MeVee, respectively. 

SANDD uses the relative charge detected in the light sensors at either end of each scintillator segment to determine the position of the event along each rod or bar (z-position). We define the charge-ratio parameter $R_{AB}$ as
\begin{equation} \label{eq_AB_ratio}
    R_{AB} = \frac{Q_{A}}{Q_{A} + Q_{B}},
\end{equation}
where $Q_A$ and $Q_B$ are the charges collected on light sensor A and B, respectively. The source position dependence on charge ratio was obtained from a series of collimated $^{137}$Cs measurements at different positions along the scintillator axis.
For events near light sensor A, $R_{AB} \rightarrow 1$. Events near the opposite end approach $R_{AB} \rightarrow 0$. In simulations, the general form of the charge ratio was found to follow an ``S'' curve over the length of each segment, especially if the attenuation length was small with respect to the length of the segment. For apparent attenuation lengths comparable to the length of the scintillator, a linear profile was found to be an adequate approximation. The linear fit was used to obtain an estimate of the position of each event along the segment
and to convert the $R_{AB}$ and sigma of $R_{AB}$ distribution of a collimated gamma-ray source into the z-position residual and its associated uncertainty. 
We did not observe significant variation in the sigma of $R_{AB}$ distribution of a collimated source taken at different positions along the scintillator axis and hence the average sigma of $R_{AB}$ distribution was used to estimate the z-position uncertainty 
throughout the entire scintillator length. Fig.~\ref{fig:zPos} shows the z-position residual and its uncertainty 
as a function of energy obtained from a collimated $^{22}$Na source (produces 511~keV and 1.27~MeV gamma-rays). We observe z-position uncertainties of approximately 2~cm, 3~cm, and 2.5~cm at 1~MeVee for component 1, 2, and 3, respectively.

\section{The simulated performance of SANDD}

In this section, the selection of positron and neutron capture candidates and the rejection of common backgrounds in the full SANDD, consisting of a full assembly of components 1, 2 and 3 (Fig.~\ref{fig_SANDD_diagram}), is discussed. Neutron capture and positron candidates were selected by establishing cuts on the energy, rod multiplicity, and PSP. Coincidence time window was then optimized to select positron-neutron pairs and to estimate the antineutrino detection efficiency. Lastly, directional capability of SANDD is discussed.

In the simulation, SANDD was surrounded by a 1.5cm-thick polyethylene layer around the sides of the scintillator (not the top and bottom where the light sensors are mounted) to reflect neutrons back into the sensitive detector volume and increase the neutron capture efficiency. The polyethylene thickness was constrained by the size of the dark box available. The neutrons and positrons that result from IBD interactions were independently simulated in SANDD. 1-MeV electrons were simulated uniformly in the scintillator segments to convert the photon hits into MeVee. A 0.2-MeVee trigger threshold was assumed. 

\subsection{Neutron capture detection efficiency}

The neutron capture efficiency in SANDD was found to be 45.4\% with 87.2\% of the neutrons were captured by $^6$Li and 12.8\% of the neutrons were capture by hydrogen. The remaining neutrons either exited SANDD (51.9\%) or were captured by hydrogen in the polyethylene layer (2.7\%). The average neutron capture time was 37.2~{\textmu}s and the average distance travelled by neutron was 7~cm. The thickness of the polyethylene layer largely affects the capture efficiency of the neutrons. For example, the neutron capture efficiency in SANDD increases to 55.5\% with 87.3\% of the neutrons were captured by $^6$Li and 12.7\% of the neutrons were captured by hydrogen when 30-cm thick polyethylene is used.

To determine the simulated detection efficiency due to the energy cut, Gaussian profiles were fitted to each component's measured energy response to neutron capture-like events (Fig.~\ref{fig_cap}). Here, the energy is defined as the maximum energy observed among the scintillator rods and bars. Two-sigma cuts in the energy parameter space were subsequently determined as [0.38,0.71]~MeV, [0.37,0.61]~MeV, and [0.35,0.61]~MeV for components 1, 2, and 3, respectively. These energy cuts correspond to a simulated 95.7\% detection efficiency.



\begin{figure}[ht]
\centering\includegraphics[width=1.0\linewidth]{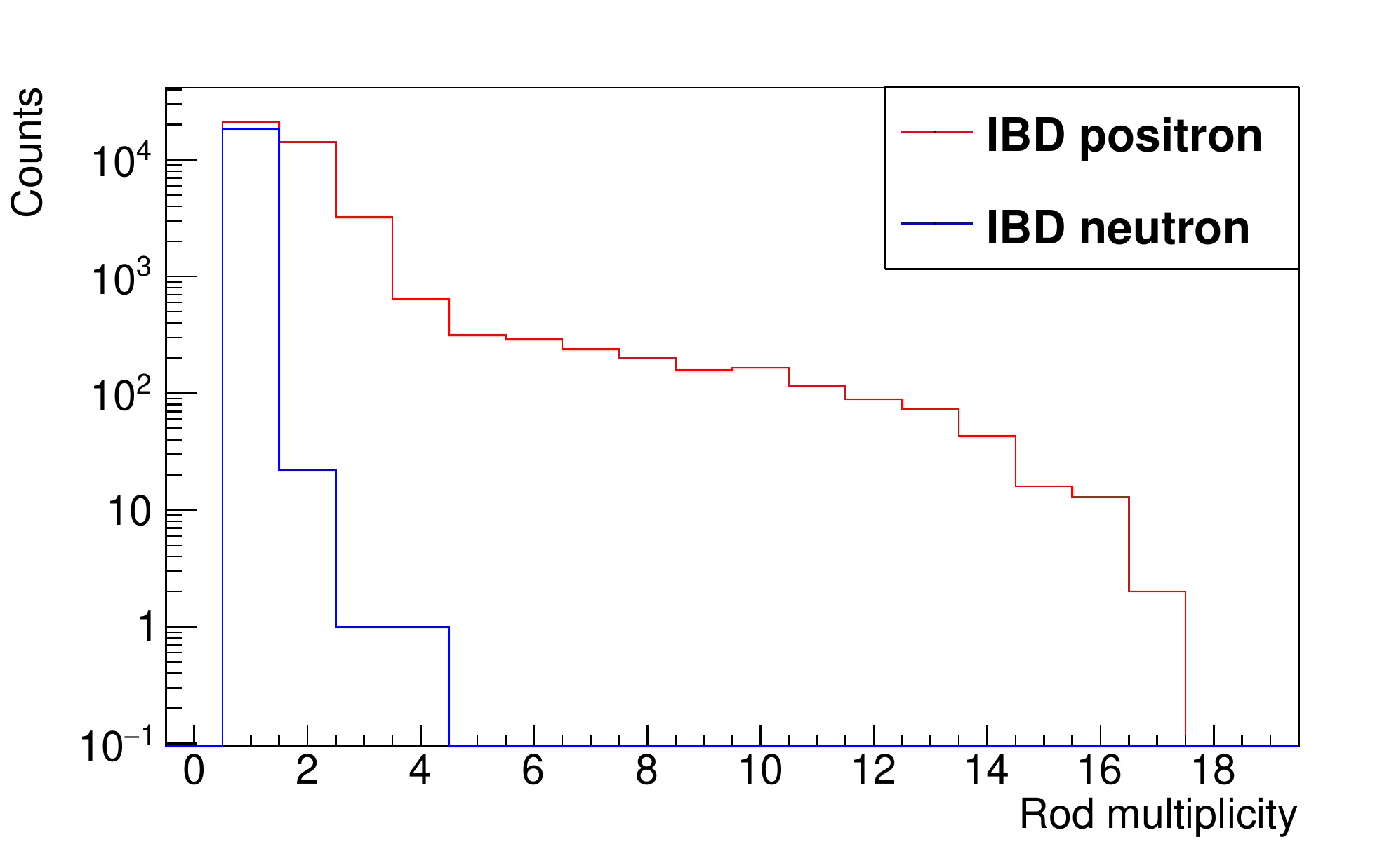}
\caption{Simulated rod multiplicity distribution of IBD-neutrons and IBD-positrons that pass the energy cuts. For IBD-neutron candidates, we require the energy to fall within the required range for neutron captures in its particular component. For IBD-positron candidates, we require the total energy (energy summed over all components) to be within [1,7]~MeVee range.}
\label{fig:gamNeuMul}
\end{figure}

Neutron capture candidates were further required to have a rod multiplicity equals to one.
Neutron captures on $^6$Li produce alpha and triton particles with a total kinetic energy of 4.78~MeV. These particles have a short range and hence their energy depositions are mostly contained within the same rod or bar~\cite{SANDD1}. Among the neutron captures that pass the two-sigma energy cut, 99.9\% have rod multiplicity equal to one (Fig.~\ref{fig:gamNeuMul}).

For neutron capture candidates, PSP was also used to differentiate between different particles. The PSP cut to identify neutron capture candidates was defined to reject background gamma rays at the 99\% level in the two-sigma energy range. 
The experimental detection efficiency due to the PSP cut was determined from the $^{252}$Cf data as shown in Fig.~\ref{fig_pspNeu}. The resulting measured neutron capture efficiencies were 73.6\%, 94.7\%, 92.7\% for components 1, 2, and 3, respectively.
Component 1 loses more lights due to the high aspect ratio, disturbing its PSP performance. 
In Fig.~\ref{fig_pspNeu}a, the PSP cut is not shown for the average central module response since a different PSP cut was applied for every rod. Multiplying the volume-weighted detection efficiency due to the PSP cut with the detection efficiencies due to the energy cut and the rod-multiplicity cut, the overall IBD-neutron detection efficiency was estimated to be
34.8\%.

\begin{figure}[ht]
\centering\includegraphics[width=1.0\linewidth]{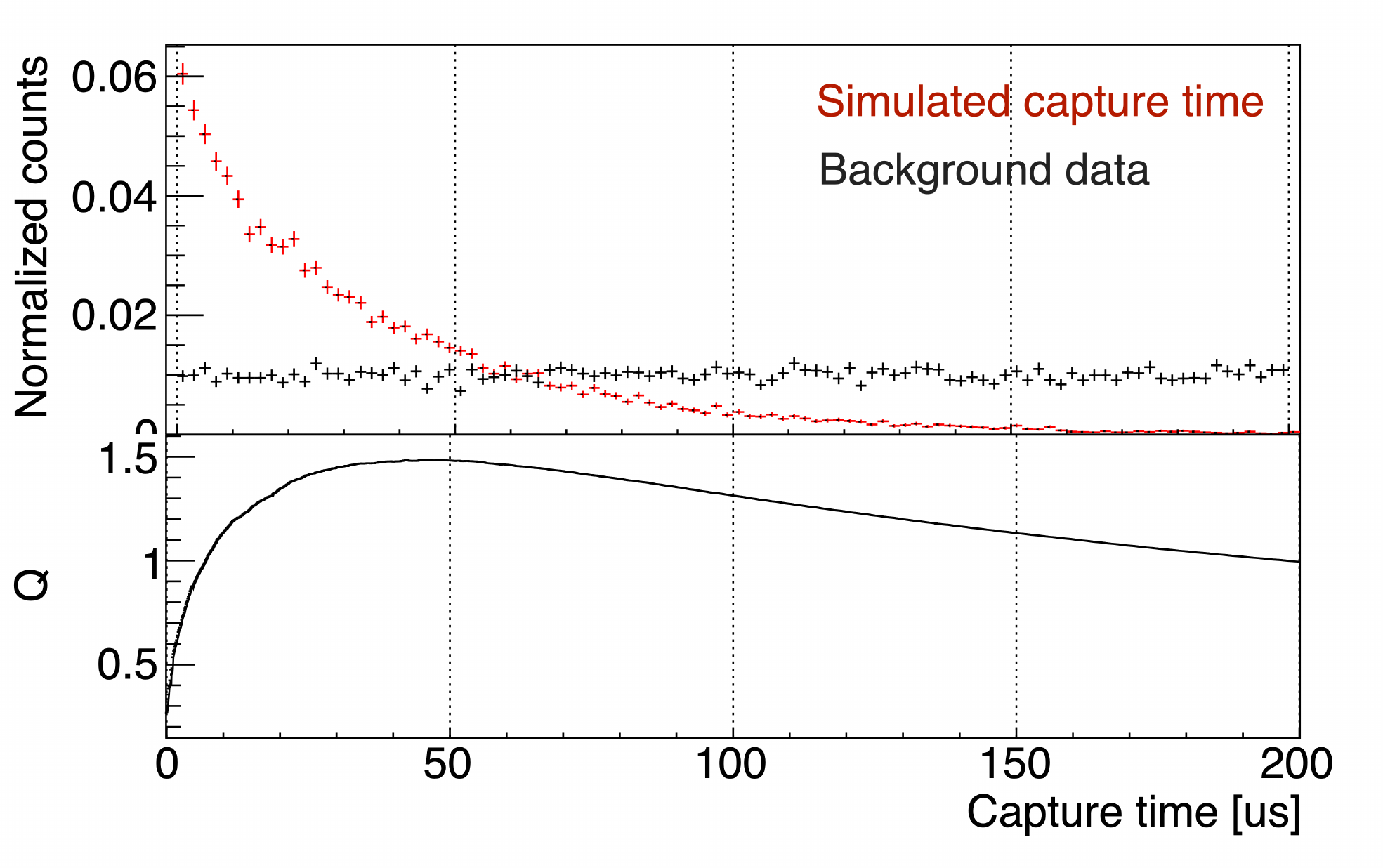}
\caption{Q analysis was applied to the measured background data and the simulated time-to-capture to obtain the best temporal cut used to isolate prompt-delayed coincident events. The average IBD neutron capture time was investigated via simulation and found to be 37.2~{\textmu}s. The optimum coincidence time cut was found to be 44~{\textmu}s.}
\label{fig:q}
\end{figure}

\subsection{Positron detection efficiency}

Positron candidates were required to deposit a minimum of 1~MeVee and a maximum of 7~MeVee summed over all the rods and bars of SANDD. 
The minimum energy requirement was chosen due to the poor pulse-shape discrimination performance below 1~MeVee (Fig.~\ref{fig_psp}). An upper energy cut of 7~MeVee was chosen to exclude the cosmic muon background. Based on the simulated energy response to IBD positrons, the [1,7]~MeVee energy requirement yielded a detection efficiency of 83.6\%.

\begin{figure*}[ht]
\begin{multicols}{3}
    \includegraphics[width=1.0\linewidth]{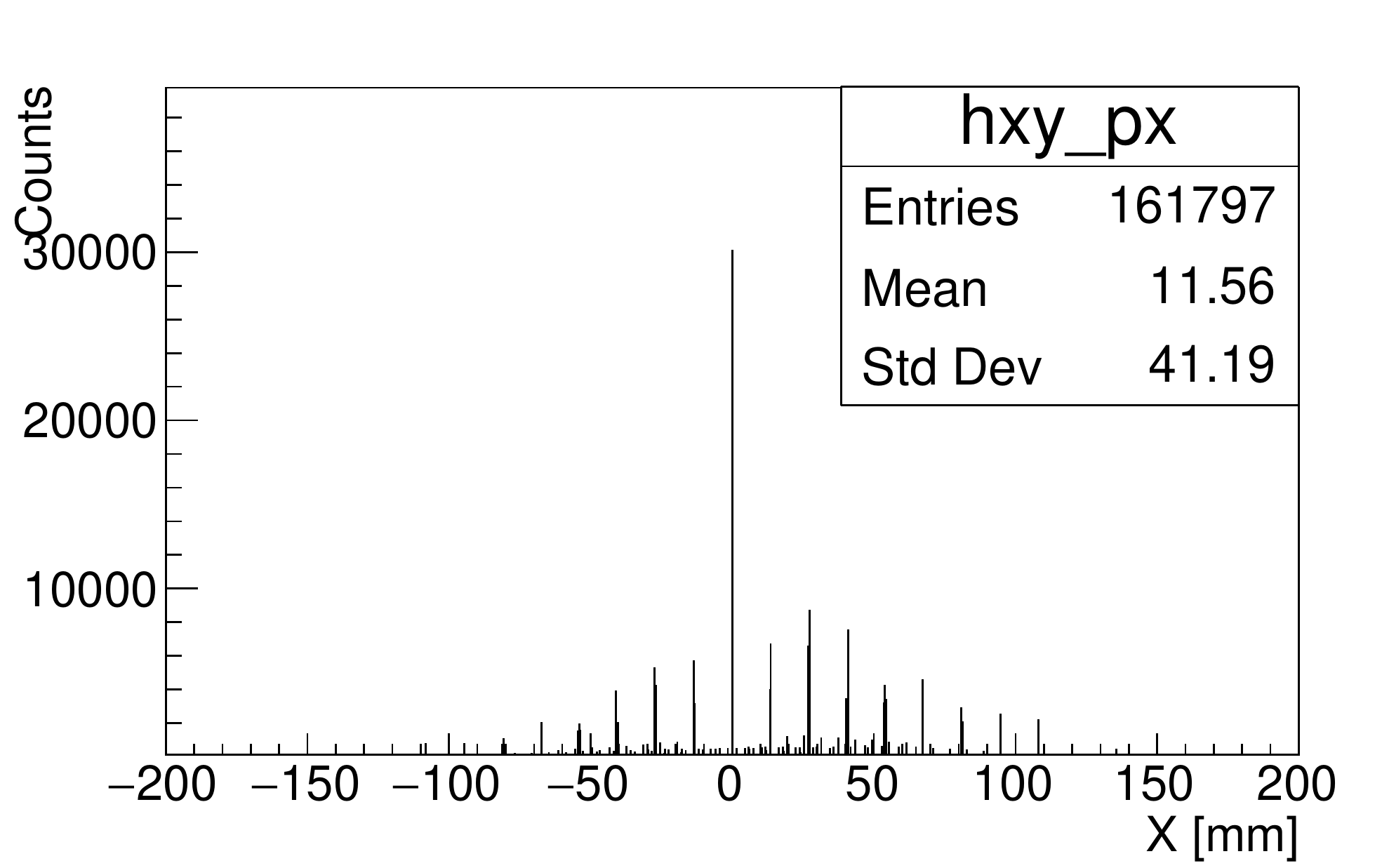} \par 
    \includegraphics[width=1.0\linewidth]{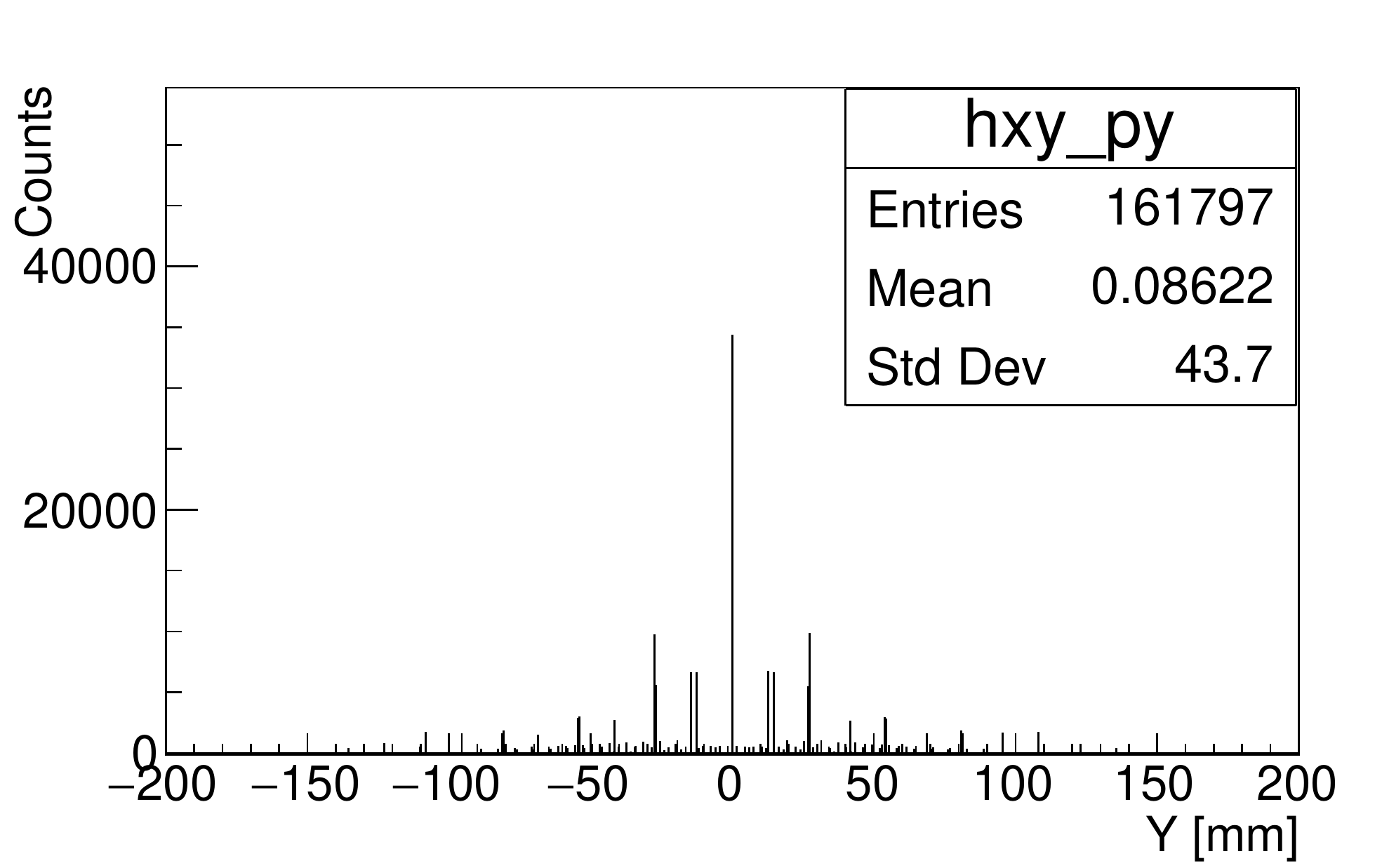} \par 
    \includegraphics[width=1.0\linewidth]{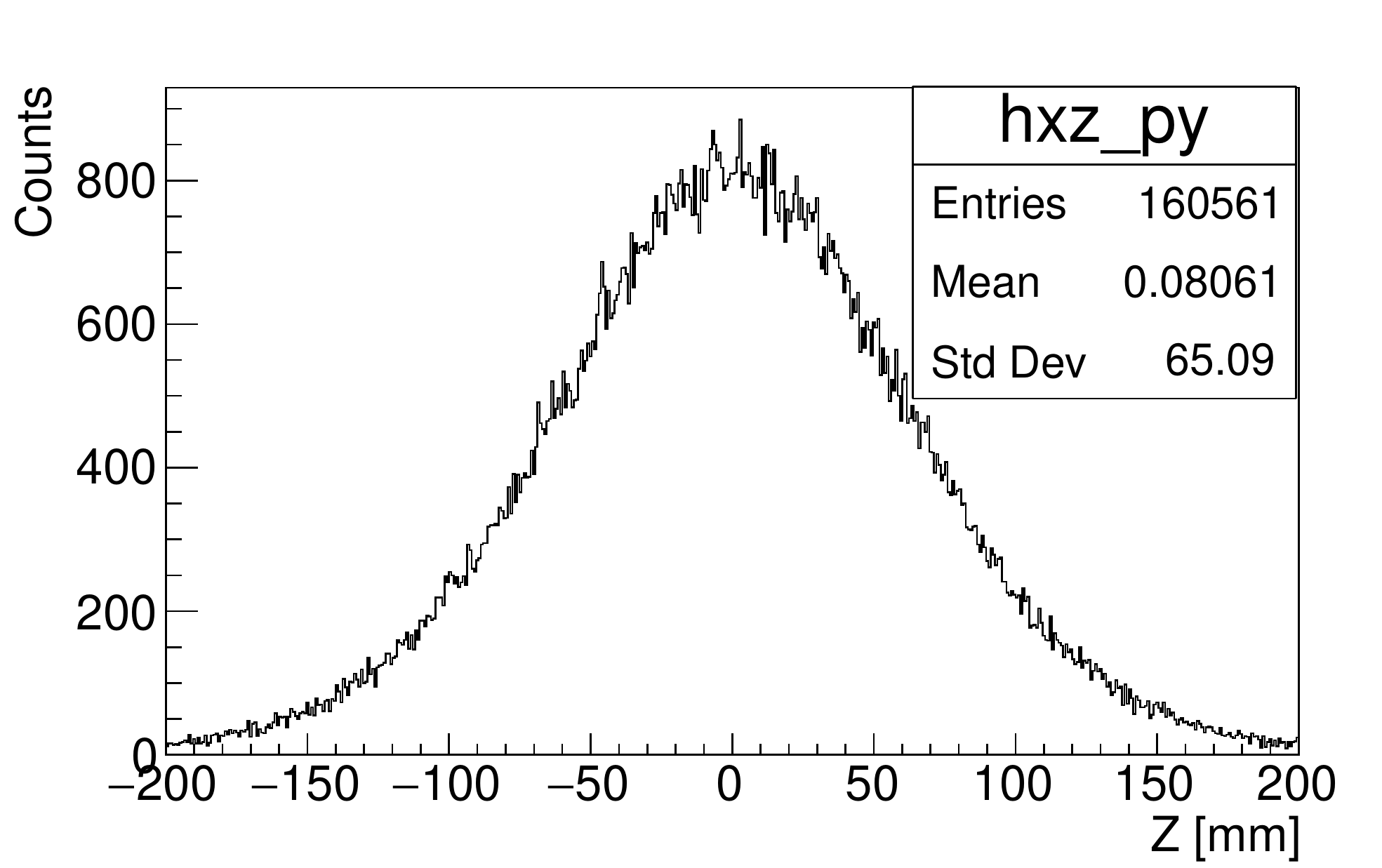} \par 
\end{multicols}
\begin{multicols}{3}
\begin{centering}
    (a) \par 
    (b) \par 
    (c) \par 
\end{centering}
\end{multicols}
\vspace{-0.3cm}
\caption{Projection of positron-neutron vectors on (a)~x-axis, (b)~y-axis, and (c)~z-axis. The simulated antineutrinos were incident along the positive x.}
\label{fig_ibdXYZ}
\end{figure*}

\begin{figure*}[ht]
\begin{multicols}{3}
    \includegraphics[width=1.0\linewidth]{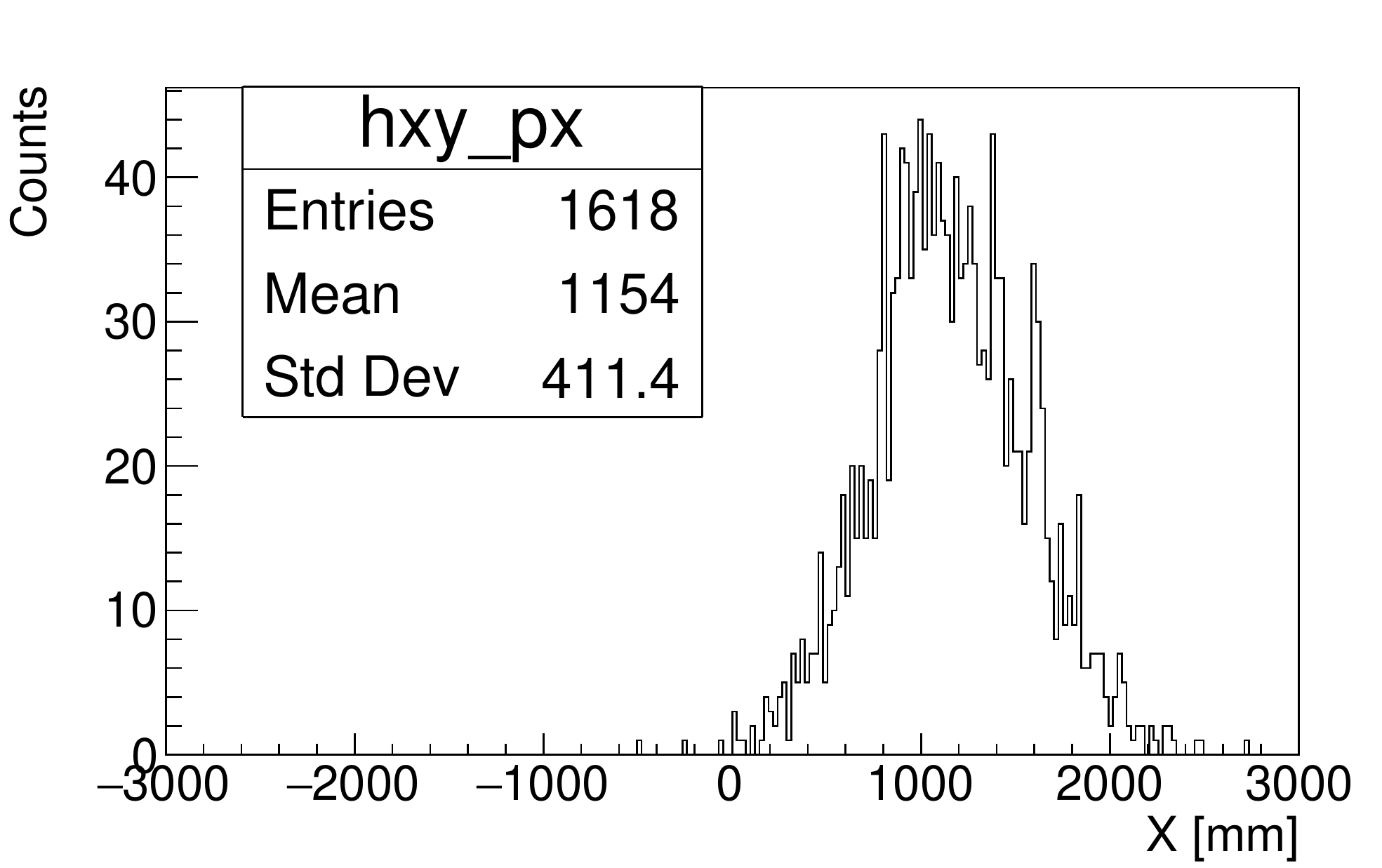} \par 
    \includegraphics[width=1.0\linewidth]{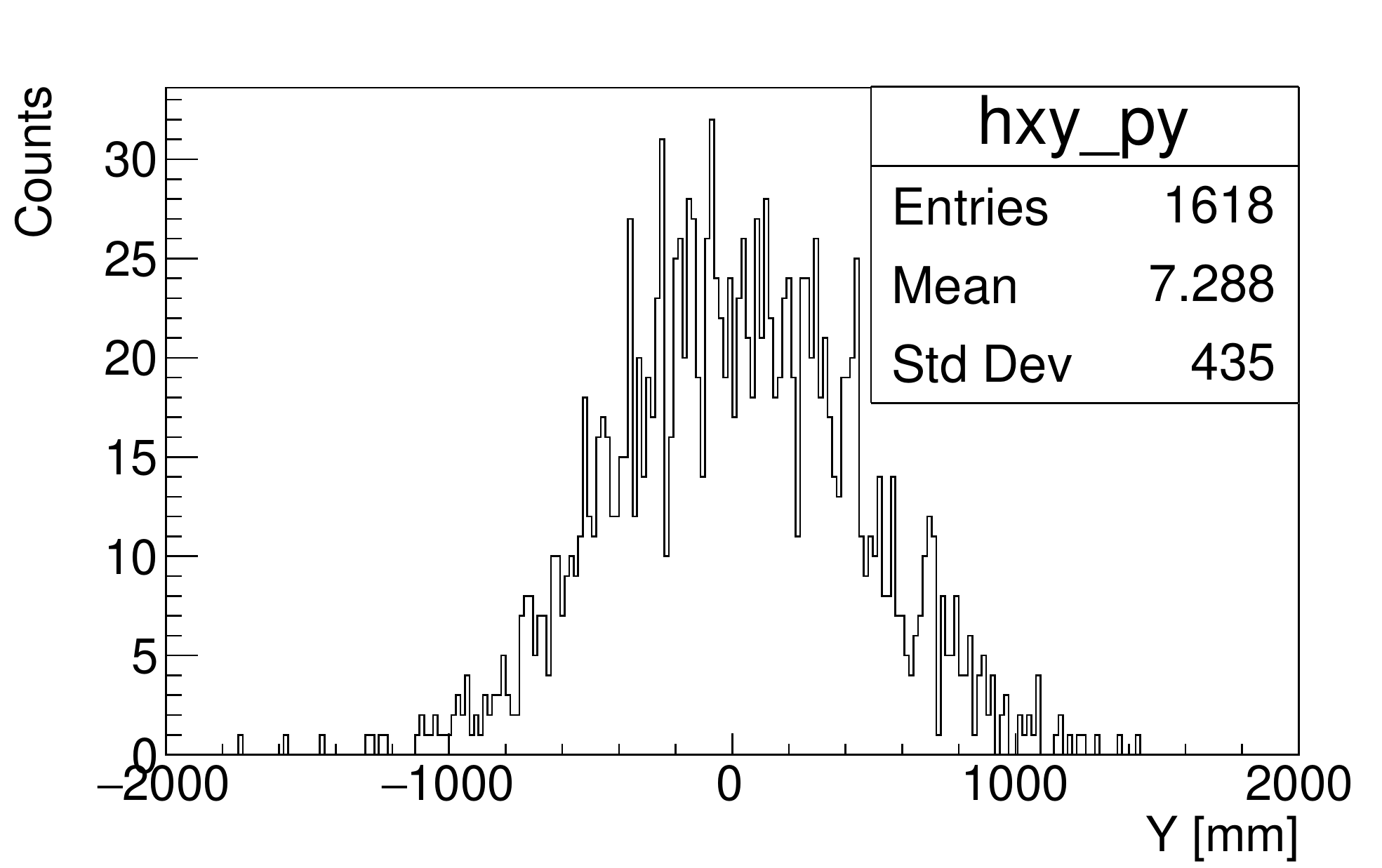} \par 
    \includegraphics[width=1.0\linewidth]{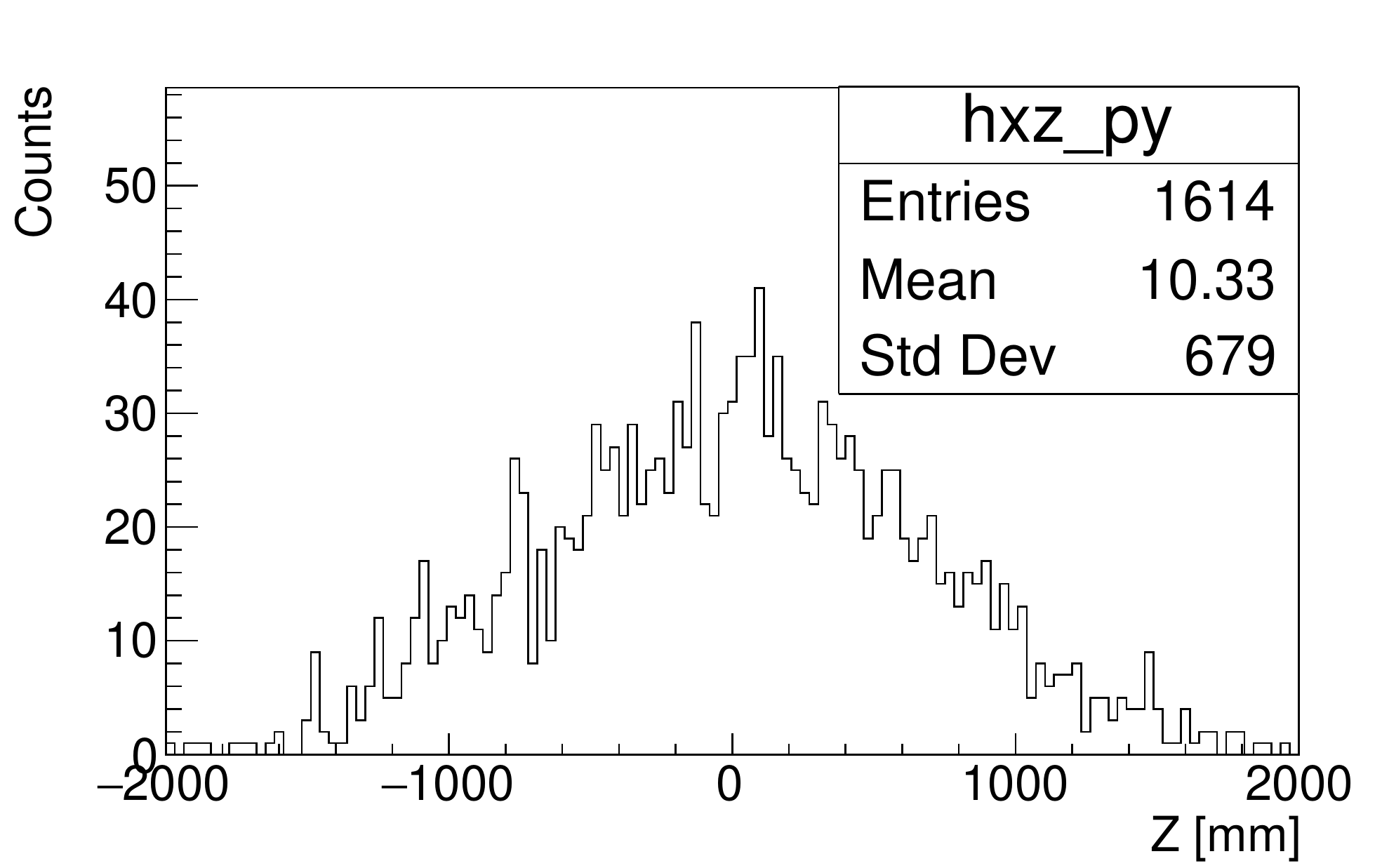} \par 
\end{multicols}
\begin{multicols}{3}
\begin{centering}
    (a) \par 
    (b) \par 
    (c) \par 
\end{centering}
\end{multicols}
\vspace{-0.3cm}
\caption{Projection of a sum of 100 positron-neutron vectors on (a)~x-axis, (b)~y-axis, and (c)~z-axis. The simulated antineutrinos were incident along the positive x.}
\label{fig_ibdXYZ_100}
\end{figure*}

\begin{table*}[ht!]
\caption{Summary of detection efficiencies} 
\label{tab_summary}
\centering
\begin{tabular}{|c||c|c|c| }
\hline
Particle & Analysis cuts & Condition & Efficiency [\%] \\
\hline
\hline
IBD neutron & Capture efficiency & & 45.4 \\
            & Fraction of capture on $^6$Li & & 87.2 \\ 
            & $\sim$0.37~MeVee$<E<\sim$0.64~MeVee  & & 95.7 \\     
            & PSP & within $E$ cut & 91.8 \\ 
            & $R_m$ $=$ 1 & within $E$ cut & 99.9 \\
\hline
            & Total IBD neutron detection efficiency & & 34.8 \\ 
\hline
\hline
IBD positron & 1~MeVee$<E<$7~MeVee &  & 83.6 \\ 
             & PSP & within $E$ cut & 95.9\\ 
\hline
            & Total IBD positron detection efficiency & & 80.2\\
\hline
\hline
  & Coincidence time $<$ 44~$\mu s$ & & 71.7 \\
\hline
\hline
 & Antineutrino detection efficiency & & 20.0 \\
\hline
\end{tabular}
\end{table*}

The PSP requirement  for positron candidates was established using a $^{252}$Cf neutron source, and setting the probability of falsely identifying a fast neutron recoil as a positron at 1\% as shown in Fig.~\ref{fig_pspGam}. The corresponding positron efficiencies due to the PSP cuts were 78.5\%, 98.2\%, and 97.0\% in the energy range [1~MeVee,7~MeVee] for components 1, 2, and 3, respectively. Multiplying the volume-weighted detection efficiency due to the PSP cut with the detection efficiencies due to the energy cut, the IBD positron detection efficiency was estimated to be 80.2\%.

\subsection{Coincidence time window}

Lastly, a temporal cut to identify positron-neutron pair candidates was established. 
A quality factor ``Q'' analysis (Eq.~(\ref{eq_q})) was applied to optimize the temporal cut used to isolate the antineutrino-like signature. The quality factor $Q$ was defined as follows:
\begin{equation}\label{eq_q}
    Q = \frac{S_c}{\sqrt{B_c}} \frac{\sqrt{B_t}}{S_t},
\end{equation}
where $S_c$ corresponds to the signal rate remaining after the cut, $B_c$ is the background rate remaining after the cut, $S_t$ is the total signal rate, and $B_t$ is the total background rate. The value of $Q$ is unity if a cut is not applied or if there is no statistical advantage gained from applying the cut. 
The neutron time-to-capture distribution (Fig.~\ref{fig:q}, top figure) was obtained via simulation (37.2~{\textmu}s average neutron capture time). 
The accidental temporal distribution (Fig.~\ref{fig:q}, top figure) was estimated from the measured background data taken with the central module where the aforementioned PSP cuts, energy cuts, and rod-multiplicity cuts were applied to the background data to estimate the time distribution between a positron-like event followed by a capture-like event.
As shown in Fig.~\ref{fig:q}, the optimum temporal cut is found to be 44~{\textmu}s, resulting in the simulated delayed coincidence detection efficiency of 71.7\%. 

\subsection{Antineutrino detection efficiency}

The antineutrino detection efficiency was calculated by multiplying the estimated neutron capture detection efficiency, the positron detection efficiency and the time coincidence efficiency (Table.~\ref{tab_summary}). 
The associated systematic uncertainties are summarized in Table.~\ref{tab_systematic}. 
To estimate the systematic uncertainties due to the variation in the simulation parameters, IBD simulations were repeated with the parameter uncertainties listed in Table.~\ref{tab_param}. For each case, we performed a simulation of 1-MeV electrons uniformly distributed in the scintilator volume to convert the photon hits into MeVee. The aforementioned analysis cuts were then applied to estimate the spread in the antineutrino detection efficiency. 
The systematic uncertainty due to the PSP cut was estimated based on the uncertainties in the Gaussian fit parameters shown in Fig.~\ref{fig_pspNeu} and Fig.~\ref{fig_pspGam}.
The antineutrino detection efficiency and its uncertainties were estimated to be 20.0\%$\pm$0.2\%(\textit{stat.})$\pm$2.1\%(\textit{syst.}).

We consider deploying SANDD near a research or power reactor.
An example of research reactor is the 85-MWt HFIR at Oak Ridge National Laboratory and an example of power reactor is the 1500-MWt Hartlepool reactor in Northern England. There are potential deployment sites located 7~m and 50~m away from the center of HFIR and Hartlepool reactor cores, which correspond to the estimated detected IBD rate of 2.6/day and 0.9/day, respectively. 


 

\begin{table}[ht!]
\caption{The relative systematic uncertainty associated with variations of the simulation parameters and PSP} 
\label{tab_systematic}
\centering
\begin{tabular}{|c||c| }
\hline
Parameters                          & Relative uncertainty \\
\hline
\hline
sigma alpha     & 3.5\% \\ 
Intrinsic attenuation length        & 3.0\% \\ 
Coupling efficiency$\times$Plastic yield  & 3.5\% \\ 
Birks constant                      & 1.5\% \\ 
$^6$Li loading                      & 8.5\% \\ 
PSP                                 & 1.0\% \\ 
\hline
\hline
Total relative uncertainty         & 10.5\% \\      
\hline
\end{tabular}
\end{table}

\subsection{Directionality}

The determination of antineutrino direction via the IBD interaction in hydrogenous media requires a probabilistic approach. The positron carries most of the energy of the IBD interaction, while the neutron carries most of the momentum. 
The IBD positron promptly annihilates, and hence its interaction position can be used as an estimate for the IBD interaction point.
The neutron undergoes a random walk before it is captured. 
Hence, on an event-by-event basis, the neutron capture location can vary considerably, and many IBD events are required to reconstruct the direction.
On average, a vector drawn from the positron to the neutron capture position will point away from the antineutrino source. 

In the simulation, the direction of the antineutrino flux was set to the SANDD local positive \textit{x} direction (see Fig.~\ref{fig_SANDD_diagram}). The aforementioned analysis cuts were applied and the 
relative positron and the thermal-neutron capture positions were estimated. 
The (\textit{x},\textit{y}) positions were estimated from the scintillator segment with the largest energy deposition.
The \textit{z} position was reconstructed based on the charge difference collected at both light sensors. 
The \textit{x,y,z}-projections of the distribution of the positron-neutron vectors are shown in Fig.~\ref{fig_ibdXYZ}.
Projections along the \textit{x}, \textit{y}, and \textit{z} axes reveal an approximately Gaussian distribution of displacements with widths 41~mm, 44~mm, and 65~mm respectively, and average displacements of 11.6~mm, 0.09~mm, and 0.08~mm.
The arrangement of the outermost layer bars of SANDD is asymmetric, resulting in the observed difference between the \textit{x} and \textit{y} distribution widths.

SANDD estimates the azimuthal angle ($\phi$) with respect to its local \textit{z} axis (Fig.~\ref{fig_SANDD_diagram}) regardless of the detector orientation with respect to the antineutrino source. Hence, the azimuthal angle was calculated based on the positron-neutron vector distribution's average shift $(\mu)$ in its local \textit{x} and \textit{y} directions:
\begin{equation} \label{eq_sandd_azimuthalAngle}
    \phi = \textrm{tan}^{-1}\left(\frac{\mu_y}{\mu_x}\right) = \textrm{tan}^{-1}\left(\frac{0.09~\textrm{mm}}{11.6~\textrm{mm}}\right) = 0.4^\circ.
\end{equation}
The reconstructed azimuthal angle was close to zero since the direction of antineutrinos was set to the local positive \textit{x}. The associated uncertainty ($\Delta\phi$) was calculated based on the average neutron displacement ($l$) and position resolution ($P$).
We calculated the average displacement as the root sum squared of the means of the \textit{x,y}-projections, yielding a value of 11.6~mm. 
We calculated the position resolution as the average of the standard deviations of the \textit{x,y}-projections, yielding a value of 42~mm.
The 1-sigma uncertainty for 100 detected IBD event was estimated as:
%
%
\begin{equation} \label{eq_sandd_aveDisplacement}
\begin{aligned}
    \Delta\phi {} = \textrm{tan}^{-1}\left(\frac{P/l}{\sqrt{N}}\right) = \textrm{tan}^{-1}\left(\frac{42.4~\textrm{mm}/11.6~\textrm{mm}}{\sqrt{100}}\right)  = 20^\circ.
\end{aligned}
\end{equation}
One could obtain similar results by repeating the calculation using the sum of 100~vectors. Figure~\ref{fig_ibdXYZ_100} shows the \textit{x,y,z}-projections of the sum of 100~vectors. The estimated average displacement and the position resolution were 1154~mm and 423~mm, yielding azimuthal angle and 1-sigma uncertainty of 0.4$^\circ$ and 20$^\circ$, respectively.
SANDD is therefore predicted to have a better directional resolution ($\pm$20$^\circ$ for 100 detected IBD events) in comparison with the Double CHOOZ experiment ($\pm$43$^\circ$ for 100 detected IBD events).

\begin{figure}[ht]
\centering\includegraphics[width=1.0\linewidth]{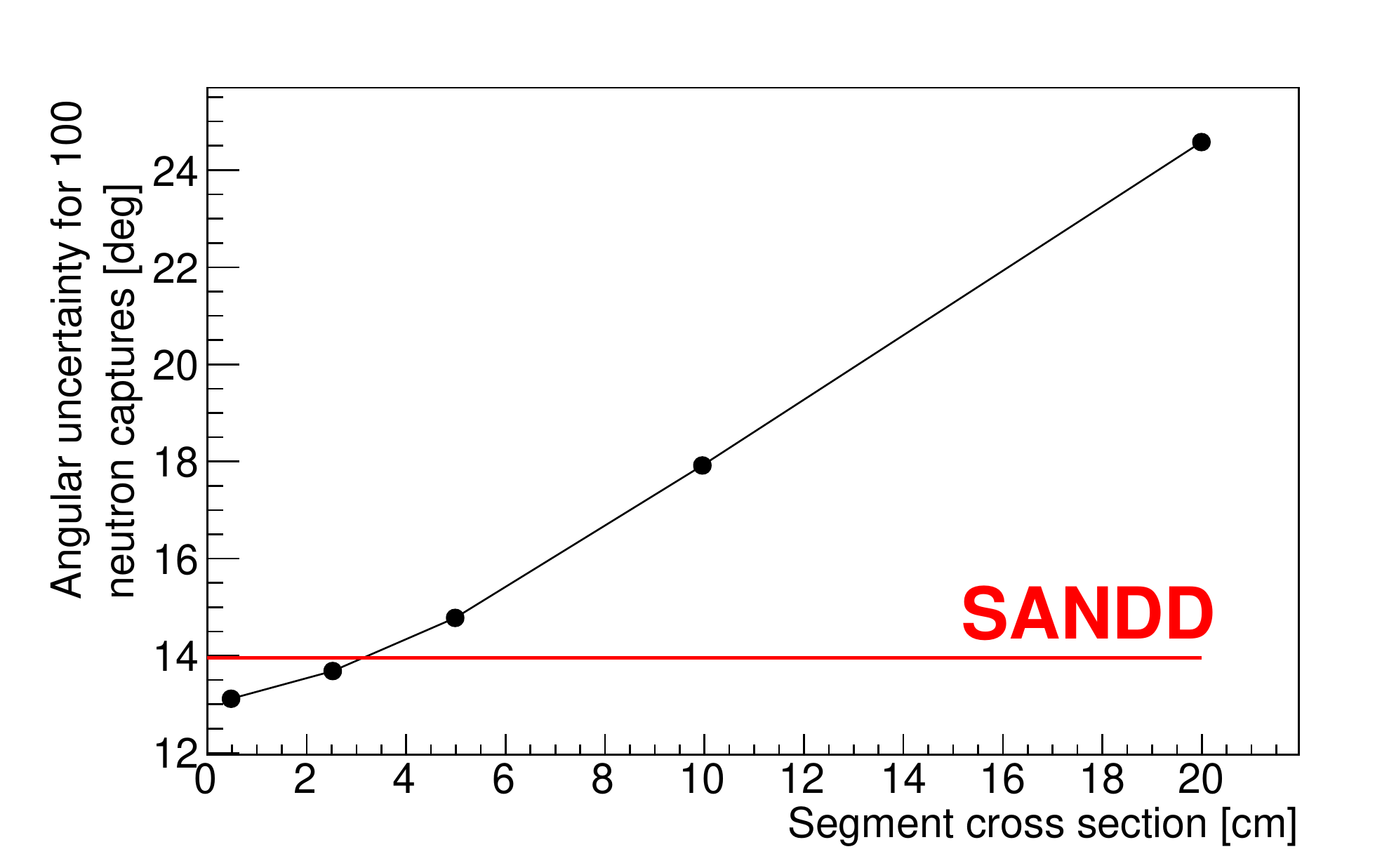}\\
(a)\\
\centering\includegraphics[width=1.0\linewidth]{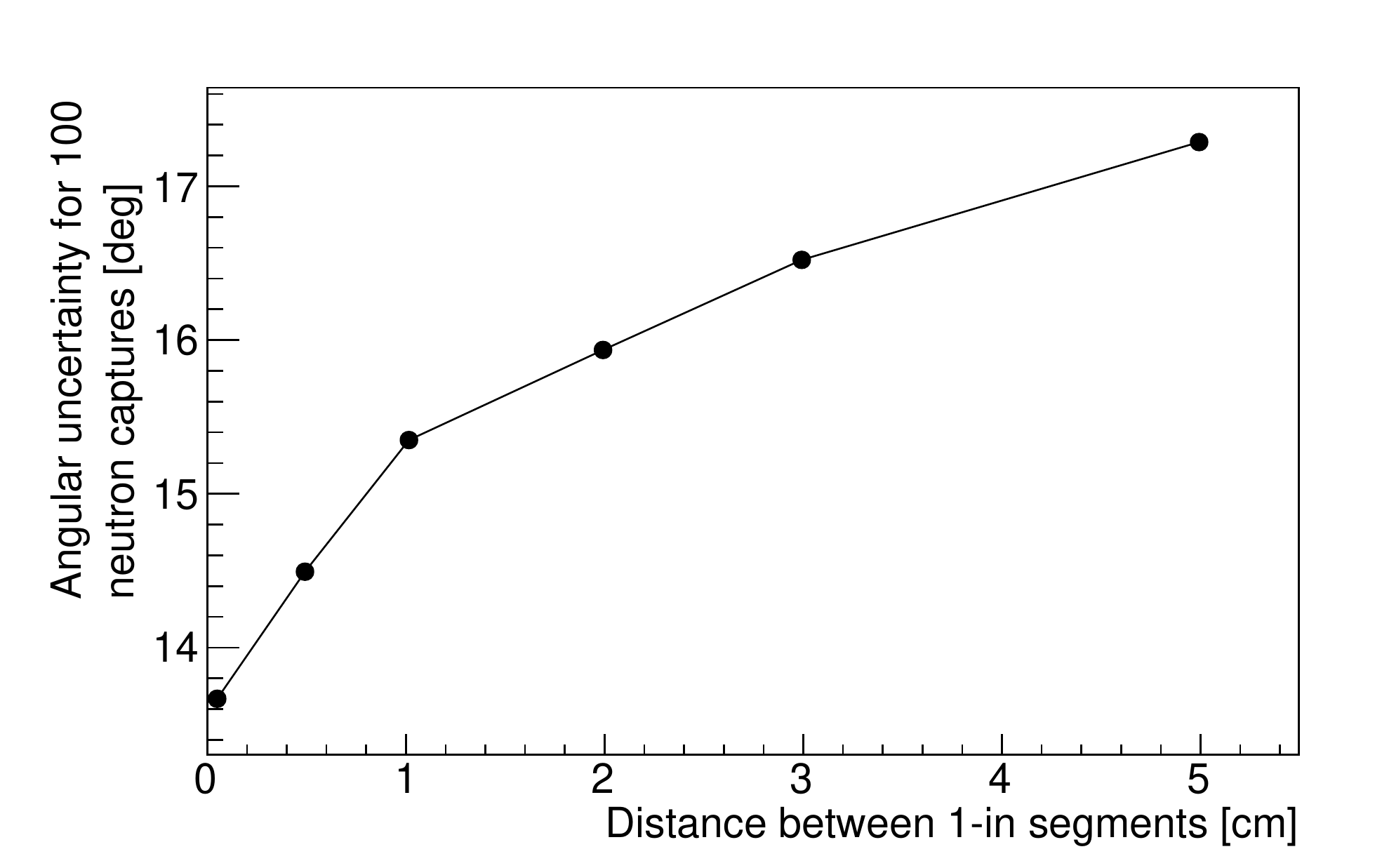}\\
(b)
\caption{Perfectly reconstructed angular uncertainty obtained from simulations of uniformly distributed IBD neutrons in a 2D array of $^6$Li-doped plastic segments with (a) varying segment cross-section and constant distance between segments at 0.6~mm and (b) varying distance between segments and constant segment cross-section at 1".}
\label{fig_toyMC}
\end{figure}

To understand how the detector design affects the angular uncertainty, we performed simple Geant4 simulations neglecting the detector response. Simulated detectors were a 2D array of $^6$Li-doped plastic segments with varying segment cross-sections and distances between segments. The simulated IBD neutrons were emitted with positive \textit{x} direction and uniformly distributed in the detector volume. Perfect position reconstruction, where neutron production and capture positions are precisely known, was assumed. This means the reconstructed (\textit{x,y}) position uncertainties correspond only to the cross-section of the segments. 
Note that in this simulation, the detector geometry and the random walk of neutrons are the only factors affecting the directional performance.
Figure~\ref{fig_toyMC}a shows perfectly reconstructed angular uncertainties for 100 IBD neutron capture events with varying segment cross-section and constant distance between segments at 0.6~mm. 
Figure~\ref{fig_toyMC}b shows the response with varying distance between segments and constant segment cross-section at 1". 
An increase in the distance between the segments results in a larger reconstructed angular uncertainty. However, when the distance between the segments approaches the cross-section of the segments, this effect becomes less significant.
A decrease in the cross-section of the segments results in a smaller reconstructed angular uncertainty. However, such improvement is less significant when the segment cross-section is less than the average 2D neutron capture distance (7.2~cm average capture distance in 3D space corresponds to 5.8~cm capture distance when projected onto \textit{x,y}-plane), suggesting that the neutron scattering plays a dominant role in determining the width of the distributions shown in Fig.~\ref{fig_ibdXYZ}.    
When we performed full simulation with the detector response considered, a detector consisting entirely of a 6$\times$6 array of component 2 bars (9.3~L detector volume) with 0.6~mm air gaps between the bars is predicted to yield an angular uncertainty of 18$^\circ$ for 100~detected IBD events. 
Such predicted angular uncertainty is smaller than that estimated for the current SANDD design (20$^\circ$ for 100~detected IBD events), which employs several larger bars and air gaps between the three different components. 
However, the current SANDD design (9~L detector volume) was chosen to provide an opportunity to investigate the particle ID performance of component 1 while retaining a good directionality and maintaining a reasonable electronics cost for a similar detector volume. 
The investigation of the component 1 performance was important to study background suppression techniques via rod multiplicity. 

\subsection{Background estimation}

We performed a cosmic neutron simulation to estimate the impact of background on the SANDD directional performance. 
Cosmic neutrons are expected to be the most dominant background aboveground~\cite{prospect2020}; they could create an IBD-like signal from a proton recoil followed by neutron capture. The Sato-Niita cosmic neutron spectrum~\cite{satoNiita} was used to sample the energy of cosmic neutrons. We used a cos$^3\theta$~\cite{preszler} angular distribution with respect to the positive \textit{y} axis to sample the direction of the cosmic neutrons since SANDD will be deployed with its \textit{x-z} plane parallel to the ground. In the simulation, the SANDD was surrounded on all sides with a 30~cm-thick polyethylene fast neutron shield. The inner 1" consisted of borated polyethylene to capture the incoming neutrons once thermalized. 
The estimated cosmic neutron-induced background detection rate was $\sim$2.9/day, suggesting that S/B $\approx$ 1 may be achievable if SANDD were to be deployed at a hypothetical 85~MWth reactor 7~m from the core (predicted signal rate is 2.6/day). For such an S/B level, the estimated azimuthal angle and the associated 1-sigma uncertainty for 100~detected IBD events are -31$^\circ$ and 31$^\circ$, respectively. This suggests that the background must be well-characterized to estimate the antineutrino direction properly.
Assuming that the background would be measurable to the same accuracy as the signal during reactor-off periods and the polyethylene shielding could reduce the rate of neutrons originating from the reactor to a negligible level during the reactor-on periods, one can estimate the antineutrino direction by subtracting the reactor-off data from the reactor-on data.
Note that we ignored background contribution from antineutrinos produced by the world rectors, neutrons produced by untagged muons, and accidental coincidences of two uncorrelated signals in our current background rate estimation. A more detailed study is needed to estimate the SANDD background level properly.

\section{Conclusion}

SANDD is a directional antineutrino detector that could potentially operate aboveground. The fine segmentation and the pulse-shape discrimination capability of SANDD allows the prompt-delayed coincidence to be discriminated against the cosmic background.
Components of SANDD were characterized and a detailed Monte Carlo simulation code was developed and tuned to investigate the performance of SANDD. 
Analysis cuts were developed to strike a balance between the antineutrino detection efficiency and the particle misclassification probability. 
The coincidence detection efficiency was estimated to be 71.7\% and the neutron and positron detection efficiencies were estimated to be 34.8\% and 80.2\%, respectively, resulting in antineutrino detection efficiency of 20\%. 
Uncertainty of $\pm$20$^\circ$ in the azimuthal direction was predicted for 100 detected antineutrino events. For comparison, Double CHOOZ experiment observes $\pm$43$^\circ$ angular uncertainty for 100 detected IBD events~\cite{caden_hawaii}.
Measurement of the direction of incoming antineutrinos could be useful for determining supernova direction, detecting geoneutrinos that could lead to better understanding of the earth's interior and its formation mechanism~\cite{tanakaWatanabe}, improving the oscillation sensitivity for reactor antineutrino experiments, and disambiguating antineutrinos emitted by multiple reactor cores in the vicinity of the detector.

While SANDD shows the promise for mobility and good directional and detection performance, this design also has several limitations.
For instance, SANDD uses a statistical approach to reconstruct the direction of antineutrino flux, which could fail in the presence of overwhelming background. 
An event-by-event directional reconstruction method using IBD has been proposed by Safdi and Suerfu~\cite{safdi2015}, where the detector volume is segmented into alternating target and capture layers with the former made sufficiently thin for neutrons to escape~\cite{safdi2015}.
Rearrangement of SANDD components could potentially allow for such technique to be used.
Another limitation is our planned active volume (9 liters) that may not allow for measurement beyond reactor on/off detection.
To monitor the core inventory of a 20--200~MWth plutonium production reactor, a ton-scale detector is required~\cite{Christensen2014,Christensen2015} to obtain reasonable statistics in the measured antineutrino spectrum.
To increase SANDD active volume while maintaining a reasonable cost of light sensors, cross section of SANDD plastic can potentially be made as large as the 2D average distance travelled by the IBD neutrons (5.8~cm). 
Further study is needed to understand the tradeoff between the antineutrino detection rate and the directional performance.

Future works include study of the cosmic background and the feasibility study of SANDD aboveground deployment.
Detector performance using different detector material such as wavelength-shifting plastic wrapped in $^6$Li-loaded ZnS scintillator may be studied.
The use of fast-timing DAQ to improve the z-position resolution and particle identification may be explored as well.

\section{Acknowledgements}

The authors wish to acknowledge Natalia Zaitseva of LLNL for helpful discussions regarding scintillator development. We thank Jingke Xu, along with  \href{http://www.ultralytics.com/}{Ultralytics} and \href{https://www.struck.de/}{Struck} engineers for their support and useful discussions. We thank Marc Bergevin for useful discussions on the Geant4 optical photon simulation. The research of F.S. was performed under the appointment to the Lawrence Livermore Graduate Scholar Program Fellowship and Rackham Merit Fellowship. The work of I.J. and F.S. was partially supported by the Consortium for Monitoring, Technology, and Verification under U.S. Department of Energy National Nuclear Security Administration award number DE-NE000863. This work was supported by the U.S. Department of Energy National Nuclear Security Administration and Lawrence Livermore National Laboratory [Contract No. DE-AC52-07NA27344, LDRD tracking number 17-ERD-016, release number LLNL-JRNL-796593].

\bibliographystyle{elsarticle-num}

\bibliography{refs_href}

\end{document}